\newcommand{\reviewera}[1]{\textcolor{black}{#1}}

\documentclass[%
 preprint,
 amsmath,amssymb,
 aps,
]{revtex4-2}
\usepackage[english]{babel} 
\usepackage{graphicx}
\graphicspath{{.},{figures/}}
\usepackage{dcolumn}
\usepackage{bm}

\usepackage{subfigure}

\usepackage{amsmath}
\usepackage{tikz}
\usepackage{tikz-cd}
\usetikzlibrary{positioning,arrows} 
\usetikzlibrary{decorations.pathreplacing} 
\usepackage{tikzsymbols} 

\usepackage{tikz}
\usetikzlibrary{shapes,arrows}
\usepgflibrary{arrows}
\usetikzlibrary{decorations.pathreplacing}
\usetikzlibrary{spy}
\usetikzlibrary{shapes,arrows}
\usetikzlibrary{calc}
\usetikzlibrary{arrows, arrows.meta}
\usepgflibrary{arrows}

\usetikzlibrary{shapes.geometric, arrows}
\tikzstyle{process} = [rectangle, minimum width=1.5cm, minimum height=1cm, text centered, text width=1.5cm, draw=black, thick, fill=black!30]
\tikzstyle{whiteprocess} = [rectangle, minimum width=1.5cm, minimum height=0.5cm, text centered, text width=2.1cm, draw=black, thick, fill=black!0]
\tikzstyle{decision} = [diamond, minimum width=4cm, minimum height=1cm, text centered, text width=3cm, draw=black, fill=black!20]
\tikzstyle{arrow} = [thick,->,>=stealth]

\usepackage{relsize}

\tikzset{fontscale/.style = {font=\relsize{#1}}
    }
    
\usepackage[export]{adjustbox}

\usepackage{scalerel}
\usepackage{stackengine,wasysym}

\begin{document}


\title[Consistent closure modeling in large eddy simulations by direct approximation of the filtered advection term]{Consistent closure modeling in large eddy simulations by direct approximation of the filtered advection term}

\author{M. Hausmann}
\email{max.hausmann@ovgu.de}
\author{B. van Wachem}
\affiliation{ 
Chair of Mechanical Process Engineering, Otto-von-Guericke-Universit{\"a}t Magdeburg, \\
  Universit{\"a}tsplatz 2, 39106 Magdeburg, Germany
}

\date{\today}

\begin{abstract}
    This article addresses the widely overlooked conceptual inconsistency of the large eddy simulation (LES) framework, namely that the commonly used advection term introduces higher wave numbers in the filtered Navier-Stokes equations than consistent with the definition of a filtered equation. It is explained why this inconsistency is the reason that flux limiters, stabilization terms, or dealiasing is often required and that the LES solution is typically mesh dependent. A consistent alternative is the direct approximation of the filtered advection term, for which we derive an expression based on an infinite series expansion with terms of increasing order in the filter width. We show that truncating the series expansion after few terms gives an expression that is highly correlated with the filtered advection term and a suitable LES model. A posteriori studies with decaying turbulence and a turbulent shear flow are conducted that reveal that the proposed approximation of the filtered advection term predicts improved kinetic energy spectra and filtered velocity correlations compared to classical LES.
\end{abstract}

\maketitle

\newpage


\section{Introduction}
Many attempts have been made over the past decades to address the long standing problem of reliable predictions of turbulent flows. One of the most promising approaches is the solution for the spatially low-pass filtered flow field, which is commonly known under the name large eddy simulations (LES). Modern applications of LES differ surprisingly little in methodology from the pioneering works of \citet{Smagorinsky1963}, \citet{Lilly1966}, \citet{Deardorff1970}, and \citet{Leonard1975}. \\
The central modeling challenge in LES consists of finding an expression for the filtered advection term, $\overline{u_i u_j}$, that contains only known filtered quantities, such as the filtered velocity $\overline{u}_i$. With very few exceptions, the filtered advection term is nowadays still approximated as $\overline{u_iu_j} = \overline{u}_i\overline{u}_j + \tau_{ij}$, where $\tau_{ij}$ are the modeled subfilter stresses, similar to the early LES studies mentioned above. There are good reasons to choose this approximation of the filtered advection term. The dyadic product of the filtered velocities, $\overline{u}_i\overline{u}_j$, can be shown to be the lowest order approximation of $\overline{u_iu_j}$ (see, e.g., \citet{Vreman1996,Leonard1997}), and constitutes a convenient choice because the advection term is identical to that in the Navier-Stokes equations (NSE) if the flow quantities are interpreted as being filtered. However, $\overline{u_iu_j}$ and $\overline{u}_i\overline{u}_j$ differ in a fundamental property that has severe consequences in practical applications of LES, that is, the range of wave numbers that both terms span. While $\overline{u_iu_j}$ is bandwidth limited at the maximum wave number of the filtering operation, $\overline{u}_i\overline{u}_j$ generally contains higher wave numbers. Therefore, it is not sufficient if the LES resolution is fine enough to resolve the filtered velocity because the high wave numbers of $\overline{u}_i\overline{u}_j$ remain unresolved. In practice, flux limiters, stabilization terms, or dealiasing are introduced to dampen the contribution of the unresolved wave numbers of $\overline{u}_i\overline{u}_j$. The discretization errors can be of the same order or even greater than the subfilter stress contribution itself \citep{Ghosal1996}. Consequently, the flow solution depends strongly on the numerical discretization, violating the fundamental principle of computational fluid dynamics that the solution should converge under mesh refinement. \\
A whole branch of LES, so-called implicit LES, relies exclusively on a manufactured discretization error without any explicit closure modeling \citep{Margolin2006,Grinstein2007}. However, the lack of a definition of a filter kernel in implicit LES renders a rigorous validation of the results with explicitly filtered DNS (FDNS) results impossible. \\
When the filter kernel is clearly defined and the non-linear interactions between the filtered and subfilter scales are explicitly accounted for by a modeled term, the LES is referred to as explicit \citep{Sagaut2005}. It is possible to derive an exact infinite series expansion for these non-linear interactions for filter kernels that are commonly used \citep{Vreman1996}. Although the terms of the series expansion have increasing orders of the filter width suggesting that they become smaller, all terms are products of spatial derivatives of the filtered velocity, which contain high wave numbers that are not suitable for LES with a coarse resolution. \citet{Germano1986} derived an exact expression for the filtered advection term based on elliptic differential filters that consists of a finite number of terms. Filtering with an elliptic differential filter is equivalent to convolution with a singular filter kernel that decays slowly in spectral space, which is not well suited to remove sufficient high wave number content to justify a coarse resolution. Another formally exact expression for the filtered advection term can be obtained by approximate deconvolution of filtered quantities \citep{Stolz1999,Stolz2001} resulting in an infinite series containing consecutively filtered quantities. Although approximate deconvolution can be conceptually suitable to approximate the filtered advection term because the approximations contains only filtered products, its convergence strongly depends on the filter kernel and the potentially large number of explicit filtering operations can add a significant computational overhead. \\
In contrast to implicit LES and LES with explicit modeling of the subfilter stress tensor, $\tau_{ij}$, that rely on flux limiters or stabilization terms, the filtered advection term $\overline{u_iu_j}$ can be modeled directly without introducing $\tau_{ij}$, making any stabilizing numerical interventions unnecessary. The direct approximation of the filtered advection term has been discussed in previous studies \citep{Lund2003,Bose2010,Singh2012}. \citet{Lund2003} makes the important point that the dyadic product of the filtered velocities used in classical LES, $\overline{u}_i\overline{u}_j$, introduces too large wave numbers. The models proposed in \citet{Lund2003,Bose2010} rely on an advection term of the type $\overline{\overline{u}_i\overline{u}_j}$ supplemented with a dynamic Smagorinsky model \citep{Germano1991,Lilly1992}. As will be demonstrated in section \ref{ssec:correlations}, $\overline{\overline{u}_i\overline{u}_j}$ is even less correlated with the filtered advection term than the dyadic product of the filtered velocities, $\overline{u}_i\overline{u}_j$, used in classical LES. The models proposed in \citet{Singh2012} contain products of filtered quantities and, therefore, introduce high wave numbers in the filtered NSE, similar to the classical advection term relying on the dyadic product of the filtered velocities. \\
In the present article, an expression for the filtered advection term is derived that is an infinite series expansion in powers of the filter width, which is formally exact and contains only terms that are filtered. Therefore, the proposed approximation for the filtered advection term only contains wave numbers consistent with the definition of a filtered quantity. It is shown that truncating the series expansion for the filtered advection term after few terms allows for an approximation of the filtered advection term that is highly correlated with the actual filtered advection term. Furthermore, it is shown that the expression predicts the total energy transfer associated to the filtered advection term and its spectral distribution increasingly accurate with increasing order. The proposed expression does not require any flux limiters, stabilization terms, or dealiasing, converges under mesh refinement, and is shown to predict more accurate flow statistics in decaying turbulence and a turbulent shear flow than classical LES. \\
The remainder of this article is structured as follows. The common practice and the limitations of classical LES are discussed in section \ref{sec:classicalLES}. In section \ref{sec:filteredadvectionterm}, the expression for the filtered advection term is derived and investigated in homogeneous isotropic turbulence (HIT). The results of two validation cases are shown and discussed in section \ref{sec:aposteriori} and guidelines for the implementation of the proposed modeling are provided in section \ref{sec:implementation} before the article concludes with section \ref{sec:conclusions}.

\section{Classical LES formulation}
\label{sec:classicalLES}
\subsection{Filtered Navier-Stokes equations}
It is assumed that the flow quantities are defined in an unconfined or infinite domain, $\Omega_\mathrm{f}$. To remove the high wave number content of a generic flow quantity, $\varPhi$, a spatial low pass filter is applied to $\varPhi$ that is defined as
\begin{align}
\label{eq:filtering}
    \overline{\varPhi}(\boldsymbol{x},t) = \int_{\Omega_\mathrm{f}}g(\boldsymbol{x}-\boldsymbol{y})\varPhi(\boldsymbol{y},t)\mathrm{d}V_y,
\end{align}
where $\overline{\varPhi}$ is the filtered flow quantity. The filter kernel, $g$, is assumed to be symmetric, spatially uniform, and satisfies
\begin{align}
    \int_{\Omega_\infty}g(\boldsymbol{x})\mathrm{d}V_x=1,
\end{align}
where $\Omega_\infty$ represents an infinite domain. Since spatially uniform filter kernels commute with spatial derivatives and filter kernels that are constant in time commute with temporal derivatives, the filtered NSE of a flow with constant density, $\rho_\mathrm{f}$, and constant dynamic viscosity, $\mu_\mathrm{f}$, are given as
\begin{align}
    \label{eq:FNSE1}
    \dfrac{\partial \overline{u}_i}{\partial x_i} &= 0, \\
    \label{eq:FNSE2}
    \rho_\mathrm{f}\dfrac{\partial \overline{u}_i}{\partial t} + \rho_\mathrm
    {f}\dfrac{\partial  \overline{u_iu_j}}{\partial x_j} &= - \dfrac{\partial \overline{p}}{\partial x_i} + \mu_\mathrm{f}\dfrac{\partial^2\overline{u}_i}{\partial x_j \partial x_j},
\end{align}
where $\overline{u}_i$ is the filtered velocity and $\overline{p}$ is the filtered pressure. If the filter kernel $g$ is zero beyond a wave number $k_\mathrm{c}$, every term in equation \eqref{eq:FNSE1} and equation \eqref{eq:FNSE2} is also zero at wave numbers larger than $k_\mathrm{c}$. Therefore, if the numerical resolution is high enough to represent functions with the wave number $k_\mathrm{c}$, it is also sufficient to resolve every term of the filtered NSE. \\
The filtered NSE as given in equations \eqref{eq:FNSE1}-\eqref{eq:FNSE2} are exact, as no modeling assumptions are introduced beyond the validity of the NSE. 

\subsection{Classical LES decomposition}
The fundamental problem of the filtered momentum equation \eqref{eq:FNSE2} is that the filtered advection term is not closed, i.e., it requires knowledge of the unfiltered velocity to be computed. Therefore, in what we refer to as classical LES, the filtered advection term is approximated as
\begin{align}
\label{eq:classicaldecomposition}
    \overline{u_iu_j} = \overline{u}_i\overline{u}_j + \tau_{ij}, 
\end{align}
where $\tau_{ij}$ is the subfilter stress tensor that requires modeling. A Taylor series expansion of the filtered velocity shows that $\overline{u}_i\overline{u}_j$ is the lowest order approximation of $\overline{u_iu_j}$ (see, e.g., \citet{Vreman1996,Leonard1997}). Furthermore, the decomposition as defined in equation \eqref{eq:classicaldecomposition} satisfies Galilean invariance exactly if $\tau_{ij}$ is Galilean invariant and the filtered NSE can be solved with a standard flow solver, as it used for DNS, by just adding an additional momentum source representing the modeled contribution of $\tau_{ij}$. 

\subsection{Limitations of classical LES formulations}
\label{ssec:limitationclassicalLES}
\begin{figure}
    \centering
    \includegraphics[width=0.6\linewidth]{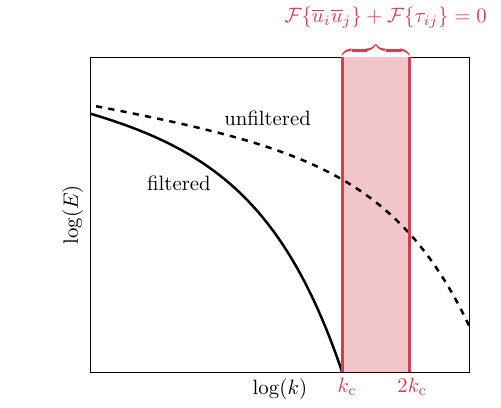}
    \caption{Sketch of the filtered and unfiltered kinetic energy spectrum, $E$, as a function of the wave number, $k$. With the classical decomposition of the advection term, the advection term and $\tau_{ij}$ are even nonzero in the highlighted wave number range that lies beyond the maximum wave number of the velocity field, $k_\mathrm{c}$. }
    \label{fig:sketchspectrum}
\end{figure}

The principle limitation of the classical LES decomposition as given in equation \eqref{eq:classicaldecomposition} is that both $\overline{u}_i\overline{u}_j$ and $\tau_{ij}$ contain nonzero amplitudes of wave numbers larger than $k_\mathrm{c}$. As a consequence, $\overline{u}_i\overline{u}_j$ and $\tau_{ij}$ require a finer resolution to be represented accurately than $\overline{u}_i$. The reason is that any product of two functions that contain wave numbers up to $k_\mathrm{c}$, generally contains wave numbers up to $2k_\mathrm{c}$. A simple example of this property is, for instance, the square of a sine-function, which gives $\sin (k_\mathrm{c}x)\sin (k_\mathrm{c}x)=\dfrac{1}{2}(1-\cos (2k_\mathrm{c}x))$. \\
After applying the Fourier transform to the filtered NSE that contain the classical LES decomposition given in equation \eqref{eq:classicaldecomposition}, it can be observed that there are only two nonzero terms remaining in the wave number range $k\in(k_\mathrm{c},2k_\mathrm{c})$ and the Fourier transformed filtered NSE reduce to
\begin{align}
\label{eq:fouriertransformFNSE}
    \mathcal{F}\{\overline{u}_i\overline{u}_j\}(\boldsymbol{k})+\mathcal{F}\{\tau_{ij}\}(\boldsymbol{k})=0, \quad |\boldsymbol{k}|\in(k_\mathrm{c},2k_\mathrm{c}),
\end{align}
where $\mathcal{F}$ indicates the Fourier transform. All other terms in the filtered momentum equation are filtered and are zero at wave numbers larger than $k_\mathrm{c}$. This is illustrated in figure \ref{fig:sketchspectrum}, where the filtered and unfiltered kinetic energy spectrum are shown together with the wave number range in which equation \eqref{eq:fouriertransformFNSE} holds. The filtered NSE with the classical LES decomposition of the advection term is still exact if the model for $\tau_{ij}$ is exact. Since the model for $\tau_{ij}$ generally contains a modeling error, equation \eqref{eq:fouriertransformFNSE} is not exactly satisfied, which results in a predicted velocity that is nonzero when solving the filtered NSE, even at wave numbers larger than $k_\mathrm{c}$. Thus, the predicted field of $\overline{u}_i\overline{u}_j$ contains wave numbers that even exceed $2k_\mathrm{c}$, which means that there is a successive propagation of the modeling error towards smaller scales. This not only violates the definition of the filtered velocity that can contain only wave numbers up to $k_\mathrm{c}$, but also requires a sufficiently high numerical resolution to represent the high wave number content accurately. \\
In practice, the numerical resolution is typically chosen to be just fine enough to resolve wave numbers up to $k_\mathrm{c}$. The larger wave numbers resulting from the inaccurate modeling of $\tau_{ij}$ cannot be resolved and appear as discretization error. Even with explicit modeling of $\tau_{ij}$, the solution strongly relies on the discretization error and therefore cannot be rigorously classified as explicit LES.\\
In addition to the active wave number range, the decomposition of the filtered advection term also has an influence on the structural properties of the filtered flow field. The filtered flow structures are directly influenced by the model for $\tau_{ij}$, but the vast majority of models relies on the Boussinesq hypothesis, i.e., on a turbulent viscosity \citep{Smagorinsky1963,Germano1991,Nicoud1999,Vreman2004}. With a turbulent viscosity model, the filtered NSE only differ by a viscosity from the unfiltered NSE, which is why classical LES with a turbulent viscosity model closely resemble the flow structures of an unfiltered DNS at a smaller Reynolds number \citep{Kamal2024,Arun2025}. However, the flow structures of the explicitly FDNS differ significantly from the LES with a turbulent viscosity model. The LES produces elongated flow structures that span a wide wave number range in different directions, but the explicit FDNS velocity contains much more isotropic filtered flow structures, which is shown in section \ref{ssec:results}.

\section{An expression for the filtered advection term}
\label{sec:filteredadvectionterm}

\subsection{General concept}
\reviewera{To close the filtered NSE given in equations \eqref{eq:FNSE1}-\eqref{eq:FNSE2}, an expression for the filtered advection term, $\overline{u_iu_j}$, is required. As explained in the previous section, the expression for the filtered advection term should be bandwidth limited with the largest wave number being the cutoff wave number determined by the filter kernel in order to ensure that the filtered velocity also remains bandwidth limited in accordance to its definition when the filtered NSE are solved. \\
In order to avoid the generation of inconsistently large wave numbers, it is required that every product of filtered quantities that is used to approximate $\overline{u_iu_j}$ must be explicitly filtered. Various defiltering or deconvolution approaches have been developed in the past (see, e.g., \citet{vanCittert1931,Geurts1997,Stolz1999,Stolz2001}) to approximate the velocity field $u_i\approx u_i^\star$. The defiltered velocity, $u_i^\star$, can then be used to approximate the filtered advection term as $\overline{u_iu_j}\approx \overline{u_i^\star u_j^\star}$. Note that the reconstruction of the velocity is never exact because the sampling of the filtered velocity on a coarse LES mesh is always associated with a loss of information. Therefore, a model for the effect of the residual velocity $u_i - u_i^\star$ is required. However, if the approximation $\overline{u_iu_j}\approx \overline{u_i^\star u_j^\star}$ is more accurate than the classical LES approximation, $\overline{u_iu_j}\approx \overline{u}_i \overline{u}_j$, less unresolved subfilter effects remain that require modeling and the residual modeling is less demanding than classical LES subgrid-scale modeling.}

\subsection{Choice of filter kernel and filter width}
\reviewera{In LES, it is very common to choose a filter kernel, $g$, with a compact support in real space, such as a top-hat filter, which is specifically convenient if the subgrid-scale model requires explicit filtering. With a top-hat filter, the filtering operation reduces to an average of the flow quantity over the neighbor mesh cells. However, a fundamental problem with this choice of the filter kernel that is often overlooked is that a top-hat filter is very ineffective in reducing the amplitudes associated to large wave numbers of flow quantities, which is the reason a filtering operation is introduced in the first place. According to the uncertainty principle \citep{Stein2003}, a nonzero function cannot be compactly supported in real and in spectral space and a rapidly decaying kernel in real space leads to a slowly decaying kernel in spectral space and vice versa. The top-hat filter kernel has compact support in real space but decays slowly in spectral space. Consequently, the filtered flow field still possesses significant energy at flow scale smaller than the cutoff of the computational mesh and large aliasing and discretization errors cannot be avoided. \\
A function that is an ideal trade-off because it is a similar function in real and in spectral space is a Gaussian. Therefore, in the following, a Gaussian filter kernel with a standard deviation $\sigma$ is used. One advantage of a Gaussian over other common filter kernels is that it decays rapidly in real and spectral space. A rapidly decaying kernel in real space allows for computationally feasible explicit filtering in LES and a kernel that is rapidly decaying in spectral space is required to remove high wave number content of the flow field to justify a coarse resolution in LES. Although a Gaussian filter kernel does not have compact support in spectral space, it decays exponentially with increasing wave number. \\
In addition to the choice of the filter kernel, it is crucial to choose the correct filter width (standard deviation of the Gaussian) with respect to the LES mesh spacing. The decisive criterion for an explicit LES is that the amplitudes of the flow quantities at the scale of the cutoff wave number introduced by the LES mesh, $k_\mathrm{c}=\pi/\Delta x$, are so small, that aliasing and discretization errors are negligible, i.e., $\hat{\overline{\varPhi}}(|\boldsymbol{k}|=k_\mathrm{c})\ll \hat{\varPhi}(|\boldsymbol{k}|=k_\mathrm{c})$. \\
Filtering in spectral space reduces to the product $\hat{\overline{\varPhi}}(\boldsymbol{k}) = \hat{g}(\boldsymbol{k})\hat{\varPhi}(\boldsymbol{k})$, where the Fourier transform of a Gaussian is given as $\hat{g}(\boldsymbol{k}) = \exp{(-|\boldsymbol{k}|^2\sigma^2/2)}$. If the filter width is chosen as $\sigma=\Delta x$, $\hat{g}(|\boldsymbol{k}|=k_\mathrm{c}) = \exp{(-\pi^2/2)}\approx 0.00719$. Therefore, filtered flow scales that are cut off by the sampling on the finite mesh have less than one percent of the amplitude of the unfiltered flow quantities. For comparison, choosing  $\sigma=\Delta x/2$ leads to $\hat{g}(|\boldsymbol{k}|=k_\mathrm{c}) = \exp{(-\pi^2/8)}\approx 0.2912$, which is associated to significant aliasing and discretization errors and the LES cannot be classified as explicit LES. Therefore, we argue that a suitable choice for the filter width is $\sigma=\Delta x$. As will be shown in section \ref{ssec:meshrefinement}, the LES with the proposed approximation of the filtered advection term is practically converged with $\sigma=\Delta x$.} 


\subsection{Differential defiltering}
\label{ssec:defiltering}
\reviewera{For the approximation of the filtered advection term, $\overline{u_iu_j}\approx \overline{u_i^\star u_j^\star}$, a suitable defiltering method is required to obtain $u_i^\star$. The most accurate defiltering would be an inversion of the filter in spectral space, which is exact up to the cutoff wave number of the LES mesh. However, this exact inversion requires periodic directions, which is why it is not suitable for most practical applications. \\
Alternatively, defiltering can be achieved by the approximate deconvolution method that estimates $u_i^\star$ using a sequence of the filtered velocity at different filter levels \citep{Stolz1999,Stolz2001}. If applied consistently with a Gaussian filter kernel and a sufficiently large filter width, the approximate deconvolution method can become computationally expensive. Defiltering can also be performed with polynomial inversion of the filter for compact filter kernels \citep{Geurts1997} but, as explained before, compactly supported filters are not suitable for LES. \\
In the present study, an alternative defiltering approach is used that is applicable for Gaussian filter kernels.} In the case of a Gaussian filter kernel, the filtering operation of a generic flow quantity, $\varPhi$, is equivalent to solving the following differential equation
\begin{align}
\label{eq:differentialfilteringphi}
    \dfrac{\partial \overline{\varPhi}(\boldsymbol{x},t,\sigma)}{\partial (\sigma^2)} = \dfrac{1}{2}\dfrac{\partial^2 \overline{\varPhi}(\boldsymbol{x},t,\sigma)}{\partial x_k \partial x_k}, \quad \overline{\varPhi}(\boldsymbol{x},t,\sigma=0) = \overline{\varPhi}|_{\sigma=0} = \varPhi.
\end{align}
For conciseness, the dependencies space, time, and filter width will be dropped in the following. \\
\reviewera{Note that precise reconstruction of $\overline{\varPhi}|_{\sigma=0} = \varPhi$ from the filtered quantity is only theoretically possible. When representing the filtered flow quantity on a finite mesh, information associated to wave numbers beyond the cutoff wave number is irretrievably lost. However, the goal of the present study is not to recover the unfiltered flow quantity, but to find an accurate approximation for $\overline{u_i u_j}$, which itself is a filtered quantity that can be represented on the same mesh that is fine enough to represent $\overline{\varPhi}$. } \\
Reformulating the convolution filtering given in equation \eqref{eq:filtering} to the solution of a differential equation alone does not give a closed expression for the filtered advection term, because the formal solution of equation \eqref{eq:differentialfilteringphi} requires knowledge of $ \overline{\varPhi}$ at smaller filter widths \citep{Johnson2020a}. Instead of considering the formal solution of equation \eqref{eq:differentialfilteringphi}, we proceed with the Taylor series expansion as a function of the squared filter width, $\sigma^2$, given as
\begin{align}
\label{eq:Tayolorseriesphi}
    \overline{\varPhi}|_{\sigma=0}  = \overline{\varPhi}|_\sigma - \sigma^2\dfrac{\partial \overline{\varPhi}}{\partial (\sigma^2)}|_\sigma + \dfrac{1}{2}\sigma^4\dfrac{\partial^2 \overline{\varPhi}}{\partial (\sigma^2)^2}|_\sigma - \dfrac{1}{6}\sigma^6\dfrac{\partial^3\overline{\varPhi}}{\partial(\sigma^2)^3}|_\sigma + \mathcal{O}(\sigma^8).
\end{align}
This allows to express the flow quantity at a filter width equal to zero as a function of the filtered flow quantity and its derivatives with respect to $\sigma^2$. To obtain an expression for the filtered advection term, a second generic flow quantity, $\varPsi$, is introduced that possesses an analog Taylor series. Computing the product $\overline{\varPhi}|_{\sigma=0}\overline{\varPsi}|_{\sigma=0}$ by multiplication of the respective Taylor series gives, after simplification,
\begin{align}
    \overline{\varPhi}|_{\sigma=0}\overline{\varPsi}|_{\sigma=0} = \overline{\varPhi}|_{\sigma}\overline{\varPsi}|_{\sigma} - \sigma^2\dfrac{\partial \overline{\varPhi}\,\overline{\varPsi}}{\partial (\sigma^2)}|_\sigma + \dfrac{1}{2}\sigma^4\dfrac{\partial^2 \overline{\varPhi}\,\overline{\varPsi}}{\partial(\sigma^2)^2}|_\sigma - \dfrac{1}{6}\sigma^6\dfrac{\partial^3 \overline{\varPhi}\,\overline{\varPsi}}{\partial(\sigma^2)^3}|_\sigma + \mathcal{O}(\sigma^8).
\end{align}
By multiplying equation \eqref{eq:differentialfilteringphi} with $\varPsi$, it can be shown that the second term of the Taylor series expansion can be expressed as a function of spatial derivatives of the filtered flow quantities
\begin{align}
    \dfrac{\partial \overline{\varPhi}\,\overline{\varPsi}}{\partial (\sigma^2)} = \dfrac{1}{2}\dfrac{\partial ^2\overline{\varPhi}\,\overline{\varPsi}}{\partial x_k \partial x_k} - \dfrac{\partial \overline{\varPhi}}{\partial x_k}\dfrac{\partial \overline{\varPsi}}{\partial x_k}.
\end{align}
Taking the derivative of the resulting term with respect to $\sigma^2$ gives an expression for the third term of the Taylor series expansion considering that filtering and spatial derivatives and spatial derivatives and derivatives with respect to $\sigma^2$ commute
\begin{align}
    \dfrac{\partial^2 \overline{\varPhi}\,\overline{\varPsi}}{\partial (\sigma^2)^2} &= \dfrac{\partial}{\partial (\sigma^2)}\left(\dfrac{1}{2}\dfrac{\partial ^2\overline{\varPhi}\,\overline{\varPsi}}{\partial x_k \partial x_k} - \dfrac{\partial \overline{\varPhi}}{\partial x_k}\dfrac{\partial \overline{\varPsi}}{\partial x_k}\right) \nonumber \\
    &=  \dfrac{\partial ^2}{\partial x_l \partial x_l} \left( \dfrac{1}{4}\dfrac{\partial ^2\overline{\varPhi}\,\overline{\varPsi}}{\partial x_k \partial x_k} -  \dfrac{\partial \overline{\varPhi}}{\partial x_k}\dfrac{\partial \overline{\varPsi}}{\partial x_k}\right) + \dfrac{\partial^2 \overline{\varPhi}}{\partial x_k \partial x_l}\dfrac{\partial^2 \overline{\varPsi}}{\partial x_k \partial x_l}.
\end{align}
Higher order terms of the Taylor series can be obtained analogously.\\
By replacing the generic flow quantities with the fluid velocity components, the filtered advection term $\overline{u_iu_j}=\overline{\overline{u}_i|_{\sigma=0}\overline{u}_j|_{\sigma=0}}$ can be expressed as a function of the filtered velocity for any arbitrary order of the filter width. An approximation of the filtered advection term up to the zeroth order term is given as
\begin{align}
    \label{eq:zerothorder}
    \overline{u_iu_j} = \overline{\overline{u}_i \overline{u}_j} + \mathcal{O}(\sigma^2),
\end{align}
up to the second order term as
\begin{align}
  \label{eq:secondorder}
  \overline{u_iu_j}= \overline{\overline{u}_i \overline{u}_j}-\sigma^2\overline{\left[\dfrac{1}{2}\dfrac{\partial^2 \overline{u}_i\overline{u}_j}{\partial x_k \partial x_k} - \dfrac{\partial \overline{u}_i}{\partial x_k}\dfrac{\partial \overline{u}_j}{\partial x_k}\right]} +  \mathcal{O}(\sigma^4),
\end{align}
and up to the fourth order term as
\begin{align}
  \label{eq:fourthorder}
  \overline{u_iu_j}&= \overline{\overline{u}_i \overline{u}_j}-\sigma^2\overline{\left[\dfrac{1}{2}\dfrac{\partial^2 \overline{u}_i\overline{u}_j}{\partial x_k \partial x_k} - \dfrac{\partial \overline{u}_i}{\partial x_k}\dfrac{\partial \overline{u}_j}{\partial x_k}\right]} \nonumber \\
  &+\dfrac{\sigma^4}{2}\overline{ \left[ \dfrac{\partial^2}{\partial x_l \partial x_l}\left( \dfrac{1}{4}\dfrac{\partial^2 \overline{u}_i \overline{u}_j}{\partial x_k \partial x_k}  - \dfrac{\partial \overline{u}_i}{\partial x_k}\dfrac{\partial \overline{u}_j}{\partial x_k}\right) + \dfrac{\partial^2 \overline{u}_i}{\partial x_k \partial x_l}\dfrac{\partial^2 \overline{u}_j}{\partial x_k \partial x_l} \right] } + \mathcal{O}(\sigma^6).
\end{align}
\reviewera{Note that there are similarities with the expression derived by \citet{Berselli2003}, who use an approximation of the filter kernel to approximate the defiltered velocity. However, the most important practical difference is that \citet{Berselli2003} derive an expression for the subfilter stresses that contains products of the filtered velocity and its derivatives, which is, as explained in sections \ref{ssec:limitationclassicalLES}, not suitable for LES.}\\
It is important to note that every single term in the proposed approximations for the filtered advection term is an explicitly filtered term. This is crucial in order to ensure that the approximations of the filtered advection term decay as fast as the Gaussian filter kernel (times a real finite constant) in spectral space, which is shown in appendix \ref{ap:spectraldecay}. The dyadic product of the filtered velocities used in classical LES, $\overline{u}_i \overline{u}_j$, for instance, decays slower than a Gaussian (times a real finite constant) in spectral space. As explained in section \ref{ssec:limitationclassicalLES} for the more intuitive case of a filter kernel with compact support in spectral space, this can cause the violation of the definition of the filtered velocity if the model for $\tau_{ij}$ is not exact. 

\subsection{Correlation with the filtered advection term}
\label{ssec:correlations}
\begin{figure}
    \centering
    \includegraphics[width=1\linewidth]{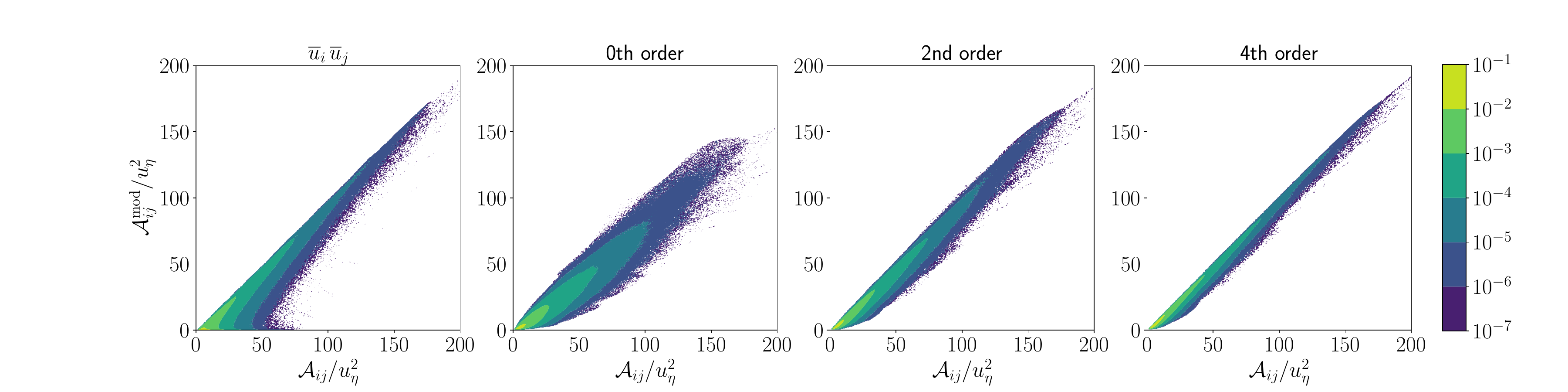}
    \caption{Joint PDF between the diagonal components ($i=j$) of different modeled filtered advection terms and the explicitly filtered advection term at a filter width of $\sigma/\Delta x_\mathrm{DNS}=4$.}
    \label{fig:jointpdfuu4}
\end{figure}

\begin{figure}
    \centering
    \includegraphics[width=1\linewidth]{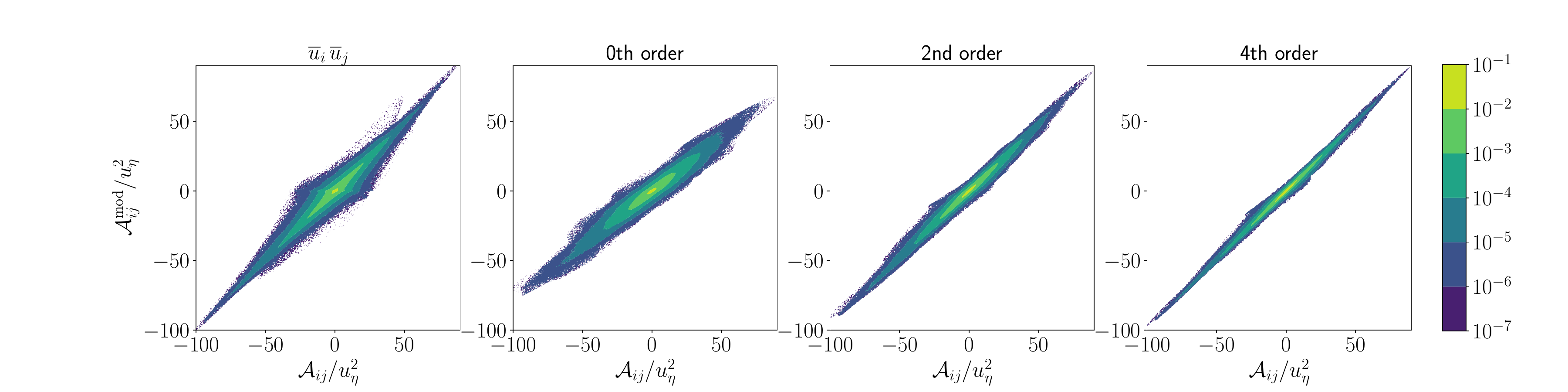}
    \caption{Joint PDF between the off-diagonal components ($i\ne j$) of different modeled filtered advection terms and the explicitly filtered advection term at a filter width of $\sigma/\Delta x_\mathrm{DNS}=4$.}
    \label{fig:jointpdfuv4}
\end{figure}

\begin{figure}
    \centering
    \includegraphics[width=1\linewidth]{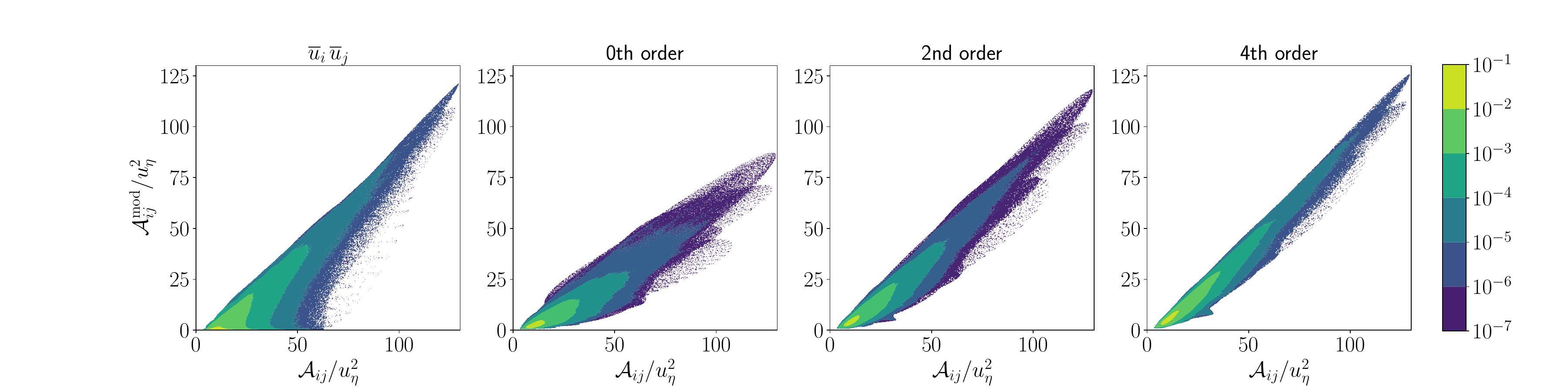}
    \caption{Joint PDF between the diagonal components ($i=j$) of different modeled filtered advection terms and the explicitly filtered advection term at a filter width of $\sigma/\Delta x_\mathrm{DNS}=8$.}
    \label{fig:jointpdfuu8}
\end{figure}

\begin{figure}
    \centering
    \includegraphics[width=1\linewidth]{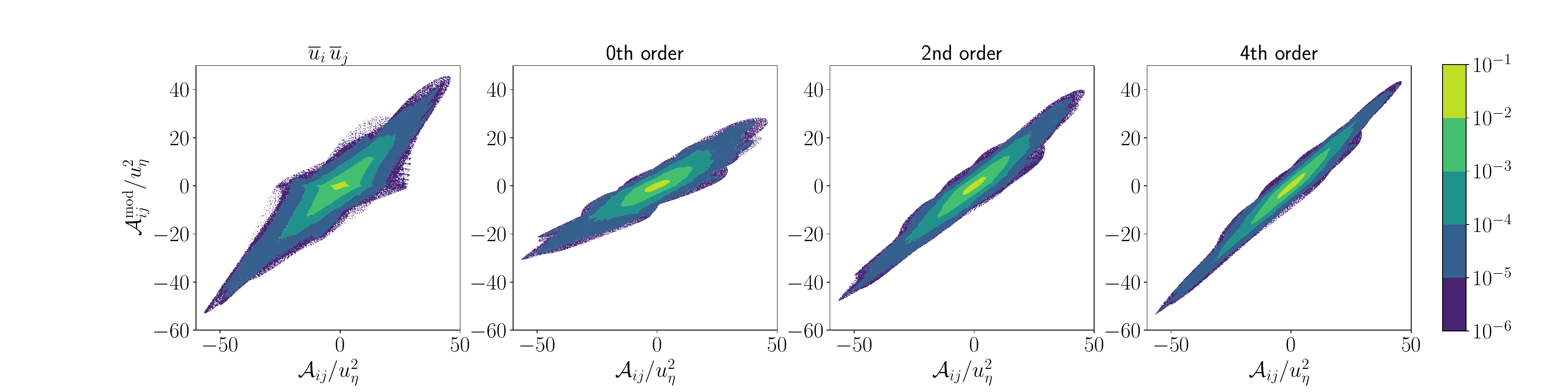}
    \caption{Joint PDF between the off-diagonal components ($i\ne j$) of different modeled filtered advection terms and the explicitly filtered advection term at a filter width of $\sigma/\Delta x_\mathrm{DNS}=8$.}
    \label{fig:jointpdfuv8}
\end{figure}

The expressions for the filtered advection term derived in the previous section become formally more accurate by including successive terms of increasing order with respect to the filter width. In this section, the accuracy of the predictions of the filtered advection term are directly evaluated for different orders based on the explicitly filtered velocity field of HIT. The velocity field is obtained from DNS of forced HIT with a Taylor-scale Reynolds number of $\mathrm{Re}_\lambda=75$. The simulations are carried out with a second order accurate finite-volume flow solver with $256^3$ mesh cells. The exact same dataset has been used in several previous studies where more details on the flow physics and the numerical solution can be found (see, e.g., \citet{Hausmann2022a,Hausmann2023,Hausmann2023a}). \\
The filtered advection term is abbreviated with
\begin{align}
    \mathcal{A}_{ij} = \overline{u_iu_j},
\end{align}
and compared to different modeled filtered advection terms, $\mathcal{A}_{ij}^\mathrm{mod}$, that are obtained by truncating the Taylor series expansion for the filtered advection term after terms proportional to the zeroth order in $\sigma$ (equation \eqref{eq:zerothorder}), after terms proportional to the second order in $\sigma$ (equation \eqref{eq:secondorder}), and after terms proportional to the fourth order in $\sigma$ (equation \eqref{eq:fourthorder}), respectively. \reviewera{When computing the expressions for the modeled filtered advection terms, a spectrally sharp filter at the cutoff wave number of the LES mesh is applied in addition to the Gaussian filter to mimic the loss of information when representing the filtered flow quantities on an LES mesh.}\\
Figure \ref{fig:jointpdfuu4} shows the diagonal and figure \ref{fig:jointpdfuv4} the off-diagonal components of the joint probability density functions (PDF) between $\mathcal{A}_{ij}^\mathrm{mod}$ for different orders and $\mathcal{A}_{ij}$ obtained from explicit filtering normalized with the Kolmogorov length and velocity scale $\eta$ and $u_\eta$, respectively. The DNS velocity field is explicitly filtered with a Gaussian filter kernel with a filter width $\sigma/\Delta x_\mathrm{DNS}=4$, where $\Delta x_\mathrm{DNS}$ is the mesh spacing of the DNS. By including terms with increasing order in $\sigma$, the diagonal and off-diagonal components of $\mathcal{A}_{ij}^\mathrm{mod}$ and $\mathcal{A}_{ij}$ are increasingly correlated. For comparison, the joint PDF is also shown with the dyadic product of the filtered velocities, $\overline{u}_i\overline{u}_j$. Including terms up to the second order in $\sigma$ is sufficient to yield a filtered advection term that is significantly better correlated with the explicitly filtered advection term than the dyadic product of the filtered velocities. Similar joint PDFs are shown for a larger filter width of $\sigma/\Delta x_\mathrm{DNS}=8$ for the diagonal components in figure \ref{fig:jointpdfuu8} and for the off-diagonal components in figure \ref{fig:jointpdfuv8}. Although the correlations between $\mathcal{A}_{ij}^\mathrm{mod}$ and $\mathcal{A}_{ij}$ are lower for the larger filter width, a similar increase of the correlation can be observed by including terms of increasing order in $\sigma$. Also for the filter width $\sigma/\Delta x_\mathrm{DNS}=8$, including terms up to $\sigma^2$ is sufficient to obtain a higher correlation than with the dyadic product of the filtered velocities. \\
The zeroth order term, which possesses a weaker correlation with the filtered advection term than the dyadic product of the filtered velocities, is used in existing models that approximate the filtered advection term \citep{Lund2003,Bose2010,Singh2012}. It may be argued that in these models and in classical LES the advection term is not applied alone but in conjunction with a modeled $\tau_{ij}$. However, in the vast majority of LES applications, the dyadic product of the filtered velocities, $\overline{u}_i \overline{u}_j$, is supplemented with a turbulent viscosity contribution, i.e., a term proportional to the strain-rate tensor, which is known to be poorly correlated with $\tau_{ij}$ \citep{Liu1994b,Borue1998,Horiuti2003}. 

\subsection{Energy transfer}
\label{ssec:energytransfer}
\begin{figure}
    \centering
    \includegraphics[width=0.9\linewidth]{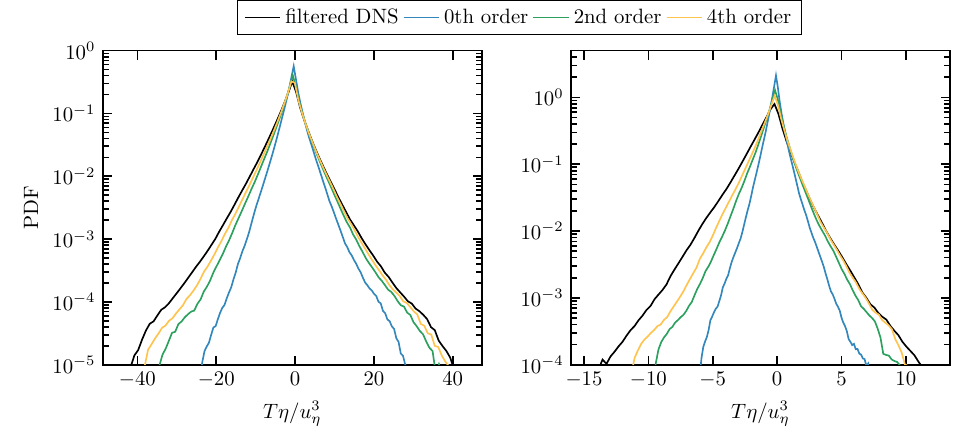}
    \caption{PDF of the energy transfer of the filtered advection term in forced HIT for the filter widths $\sigma/\Delta x_\mathrm{DNS}=4$ (left) and $\sigma/\Delta x_\mathrm{DNS}=8$ (right).}
    \label{fig:PDFsenergytransfer}
\end{figure}

\begin{figure}
    \centering
    \includegraphics[width=0.5\linewidth]{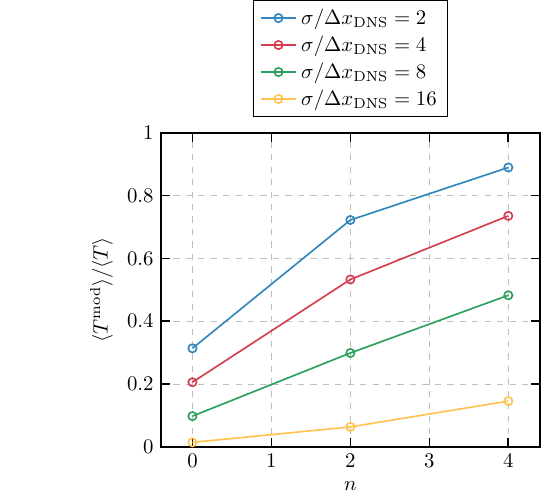}
    \caption{Fraction of the total energy transfer captured by the model for the filtered advection term for different orders $n$ and different filter widths $\sigma$.}
    \label{fig:totalenergytransfern}
\end{figure}

\begin{figure}
    \centering
    \includegraphics[width=0.5\linewidth]{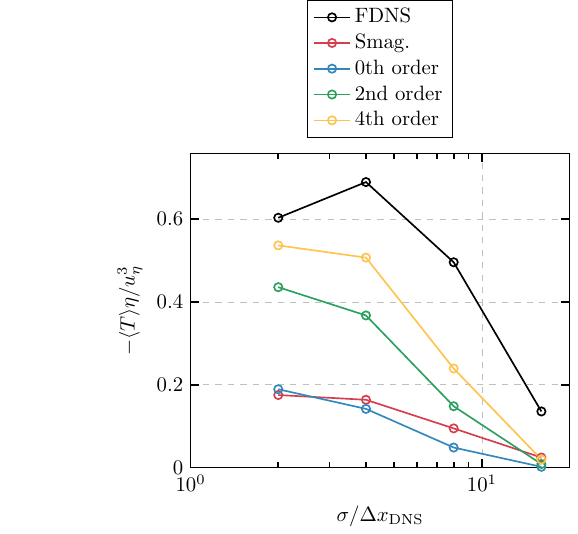}
    \caption{Total energy transfer for different approximations of the filtered advection term as a function of the filter width.}
    \label{fig:totalenergytransfersigma}
\end{figure}

\begin{figure}
    \centering
    \includegraphics[width=0.9\linewidth]{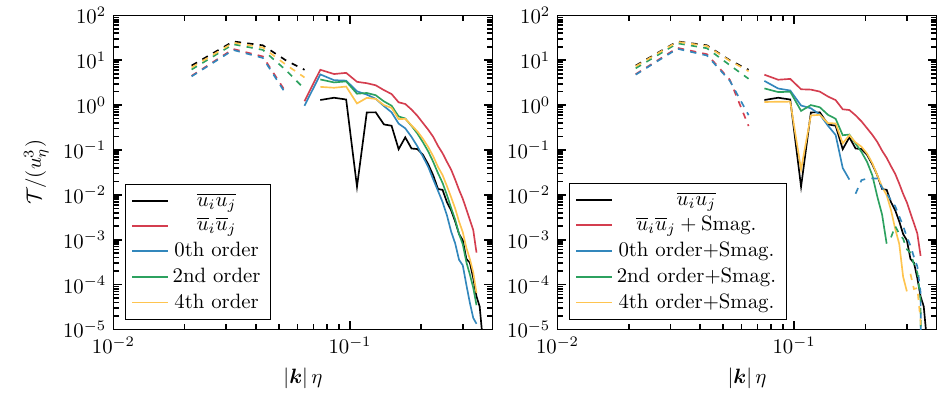}
    \caption{Energy transfer spectra for different approximations of the filtered advection term at a filter width of $\sigma/\Delta x_\mathrm{DNS}=4$.  Dashed lines indicate negative energy transfer.}
    \label{fig:energytransfer}
\end{figure}

An expression for $\mathcal{A}_{ij}^\mathrm{mod}$ that is highly correlated with $\mathcal{A}_{ij}$ is not guaranteed to be a suitable LES closure model. A crucial property of a suitable closure model is the prediction of realistic energy transfer between the filtered and subfilter scales. The energy transfer resulting from the filtered advection term is denoted as $T$ and defined as 
\begin{align}
    T = -\overline{u}_i\dfrac{\partial}{\partial x_j}\mathcal{A}_{ij}.
\end{align}
The modeled energy transfer resulting from the modeled filtered advection term, $\mathcal{A}_{ij}^\mathrm{mod}$, is denoted as $T^\mathrm{mod}$. \\
On average, the filtered advection term removes energy from the filtered scales, distinguishing it from the dyadic product of the filtered velocities used in classical LES, $\overline{u}_i\overline{u}_j$, which does not cause any net energy transfer between filtered and subfilter scales. The pointwise energy transfer of the filtered advection term, however, can be positive and negative, representing a bidirectional energy exchange between the scales. \\
\citet{Lund2003} makes the correct observation that the approximation of the filtered advection term $\overline{\overline{u}_i\overline{u}_j}$ causes a nonzero net energy transfer and this net energy transfer is referred to as “false dissipation”. However, since the filtered advection term that $\overline{\overline{u}_i\overline{u}_j}$  approximates itself produces a nonzero net energy transfer, “false dissipation” is certainly not a suitable term. The energy transfer induced by $\overline{\overline{u}_i\overline{u}_j}$ may not be very accurate, but the fact that there is net energy transfer is a desirable property. \\
Figure \ref{fig:PDFsenergytransfer} shows the PDF of the energy transfer in forced HIT of the FDNS together with the modeled advection up to terms of the zeroth, second, and fourth order for the filter widths $\sigma/\Delta x_\mathrm{DNS}=4$ and $\sigma/\Delta x_\mathrm{DNS}=8$. The PDF of the energy transfer is skewed and possesses a longer tail in the negative direction, which is associated to a removal of energy from the filtered scales. However, a significant portion of the energy transfer is positive. The positive energy transfer is caused by backscattering of energy from the subfilter scales to the filtered scales and by positive energy transfer between the filtered scales themselves. In figure \ref{fig:PDFsenergytransfer} it can be observed that the positive energy transfer is captured very well by the modeled advection term up to the fourth order, even for the larger filter width. With increasing order of the modeled advection term, the PDFs of the energy transfer approach the correct PDF from the FDNS. However, for all orders of the modeled advection term, the events associated to significant removal of kinetic energy from the filtered scales occur too rarely. \\
The total energy transfer is further analyzed by examining the fraction of the energy transfer of the filtered advection term that is captured by the modeled filtered advection term for different filter widths and different order $\sigma^n$, as shown in figure \ref{fig:totalenergytransfern}. The total energy transfer is denoted as $\langle T\rangle$, where the angular brackets indicate ensemble averaging. It is observed that the total energy that is removed from the filtered scales is predicted increasingly accurate by including terms of increasing order $n$ and decreasing filter width. Therefore, a sufficient removal of energy would require including terms of a high order $n$ for large filter widths. Although it is straightforward to derive high order expressions for the filtered advection term, high orders in $\sigma$ are associated to high order spatial derivatives of the filtered velocity, which can become expensive to evaluate and potentially less accurate if the filter width and mesh spacing are kept constant. Since the correlation of the modeled filtered advection term with the filtered advection term is still high for large filter widths, it can be beneficial in practice to supplement the model for the filtered advection term with a simple dissipative mechanism for large filter widths instead of increasing the order $n$.\\ 
In figure \ref{fig:totalenergytransfersigma}, the total energy transfer and its approximations are shown as a function of the filter width. In addition to the approximations of the filtered advection term with different orders, the energy transfer is also shown for a Smagorinsky type approximation that is given as
\begin{align}
    \label{eq:Smagorinskyterm}
    \mathcal{A}_{ij}^\mathrm{mod} = -(C_\mathrm{S}\sigma)^2|\overline{S}|2\overline{S}_{ij},
\end{align}
where $\overline{S}_{ij}$ is the filtered strain-rate tensor, $C_\mathrm{S}$ is the Smagorinsky constant that is assumed as $C_\mathrm{S}=0.2$ for the remainder of this article, and $|\overline{S}|$ is a norm of the filtered strain-rate tensor that is defined as 
\begin{align}
    |\overline{S}| = \sqrt{2\overline{S}_{ij}\overline{S}_{ij}}.
\end{align}
It is confirmed by examining the results presented in figure \ref{fig:totalenergytransfersigma} that a higher order approximation of the filtered advection term significantly improves the predicted total energy transfer, but even for $n=4$ some removal of energy from the filtered scales is missing. With the commonly used Smagorinsky approximation, an even larger discrepancy to the required energy transfer remains. In classical LES, where the advection term is the dyadic product of the filtered velocities $\overline{u}_i\overline{u}_j$, the Smagorinsky contribution is the only term of the approximation for the filtered advection term that leads to a nonzero net energy transfer. \\
\reviewera{Although the proposed approximations of the filtered advection term predict a net energy transfer from the filtered scales to the subfilter scales, the net energy transfer is generally smaller than the net energy transfer of $\overline{u_iu_j}$. With increasing order in $\sigma$ of the proposed approximations of the filtered advection term, more energy is transferred to the subfilter scales. However, even by increasing the order of the approximations further, the energy transfer is not necessarily predicted correctly because a full reconstruction of the defiltered velocity from filtered quantities represented on a finite mesh is impossible. Therefore, the proposed approximations of the filtered advection term have to be supplemented with an additional model that represents the unresolved interactions. Since the correlation of the approximation with the filtered advection term is very high, an additional model mainly has to predict the missing energy transfer, which can be realized with a Smagorinsky model or many other models. Because of the high correlation and the fact that a good proportion of the energy transfer is already captured by the approximation of the filtered advection term, an additional model can be less demanding compared to the classical advection term $\overline{u}_i\overline{u}_j$.}\\
If only the total energy transfer between filtered and subfilter scales is considered, supplementing the approximation for the filtered advection term with a Smagorinsky contribution appears to be a suitable intervention to obtain the correct energy transfer. However, to obtain a realistic filtered flow field, the energy must be removed from the correct scales, i.e., the spectral distribution of the energy transfer must be correct. The kinetic energy spectrum of the filtered velocity is altered by the energy transfer spectrum that is defined as
\begin{align}
    \mathcal{T}(k) = -\oint_{|\boldsymbol{k}|=k} \Re\{\hat{\overline{u}}_i^*(\boldsymbol{k})\widehat{\dfrac{\partial \mathcal{A}_{ij}}{\partial x_j}}(\boldsymbol{k})\} \mathrm{d}\mathcal{S}(k),
\end{align}
where $\hat{.}$ indicates the Fourier transform, $.^*$ indicates a complex conjugate, and $\Re$ refers to the real part. The integration is carried out over spherical shells in wave number space $\mathcal{S}(k)$. According to Parseval's theorem
\begin{align}
    \langle T\rangle = \int_0^\infty\mathcal{T}(k)\mathrm{d}k,
\end{align}
which means that $\mathcal{T}$ integrates to zero for the dyadic product of the filtered velocities $\overline{u}_i\overline{u}_j$. \\
Figure \ref{fig:energytransfer} shows the energy transfer spectrum with the explicitly filtered advection term, with different orders of approximation of the filtered advection term, and with the dyadic product of the filtered velocities. Furthermore, the energy transfer spectra are shown with the approximations of the filtered advection term supplemented with the Smagorinsky term given in equation \eqref{eq:Smagorinskyterm}. \\
All of the energy transfer spectra are negative for the large filtered scales and positive for the small filtered scales, which means that all approximations transfer energy from the large to the small filtered scales. Since the net energy transfer of the dyadic product of the filtered velocities, $\overline{u}_i\overline{u}_j$, is zero, the small filtered scales receive exactly the same energy as is removed from the large filtered scales. With increasing order of the proposed approximations of the filtered advection term, the energy removed from the large filtered scales increases and the energy added to the small filtered scales decreases, such that with increasing order the energy transfer spectrum approaches that of the FDNS. \\
By adding the Smagorinsky term given in equation \eqref{eq:Smagorinskyterm} to the approximations of the filtered advection term, a dissipative component is added. This additional dissipation is not sufficient for the dyadic product of the filtered velocities to predict an accurate energy transfer spectrum. The fourth order approximation of the filtered advection term together with the Smagorinsky term leads to a very accurate prediction of the energy transfer spectrum across the whole wave number range. Therefore, the Smagorinsky term not only provides the required total dissipation but also the correct spectral distribution, such that the energy transfer is correct at all wave numbers. \\
In summary, the fourth order approximation of the filtered advection term together with the Smagorinsky term can predict the correct total energy transfer and even the correct spectral distribution of the energy transfer. Furthermore, it is shown in the previous sections that the fourth order approximation of the filtered advection term is highly correlated with the filtered advection term and that it does not introduce wave numbers in the filtered NSE that exceed the maximum wavenumber consistent with the definition of a filtered quantity.

\subsection{Galilean invariance}
In addition to producing accurate flow physics, the model for the filtered advection term must also comply with the fundamental principle that the observed physics is independent of the velocity of the observer, i.e., that Galilean invariance is satisfied. Galilean invariance is satisfied if the momentum equation in a frame of reference moving with the constant velocity $\boldsymbol{u}_\mathrm{ref}$ is the same as in the fixed frame of reference. If $u_i$ is the velocity in the moving frame of reference, the velocity in the fixed frame is given as
\begin{align}
    \mathcal{U}_i = u_i(\boldsymbol{x}-\boldsymbol{u}_\mathrm{ref}t,t) + u_{\mathrm{ref},i},
\end{align}
Since the filtered advection term $\overline{u_iu_j}$ can be verified to lead to a Galilean invariant filtered momentum equation, Galilean invariance of the modeled advection term is satisfied if
\begin{align}
    \overline{\mathcal{U}_i\mathcal{U}_j}-\mathcal{A}_{ij}^{\mathrm{mod}}(\overline{\mathcal{U}}_i) = \overline{u_iu_j} - \mathcal{A}_{ij}^{\mathrm{mod}}(\overline{u}_i).
\end{align}
that is, the modeled advection term produces the same additional terms when changing the frame of reference as the filtered advection term and the error introduced by modeling the filtered advection term is the same in all frames of reference. \\
The dyadic product of the filtered velocities turns out to exactly satisfy Galilean invariance since 
\begin{align}
    \overline{\mathcal{U}_i\mathcal{U}_j}-\overline{\mathcal{U}}_i\overline{\mathcal{U}}_j = \overline{u_iu_j} - \overline{u}_i\overline{u}_j.
\end{align}
In practice, however, Galilean invariance is almost never satisfied because advection schemes with flux limiters and finite element stabilization terms typically introduce a velocity dependence of the discretization. In particular, this concerns implicit LES that inherently rely on the dissipative properties of the flux limiters or stabilization terms. \\
Checking Galilean invariance of proposed approximations of the filtered advection term gives
\begin{align}
    \label{eq:Galileaninvariance0}
    \overline{\mathcal{U}_i\mathcal{U}_j}-\mathcal{A}_{ij}^{\mathrm{mod}}(\overline{\mathcal{U}}_i) = \overline{u_iu_j} - \mathcal{A}_{ij}^{\mathrm{mod}}(\overline{u}_i) + u_{\mathrm{ref},i}\underbrace{(\overline{u}_j-\overline{\overline{u}}_j)}_{\mathcal{O}(\sigma^2)} + \underbrace{(\overline{u}_i-\overline{\overline{u}}_i)}_{\mathcal{O}(\sigma^2)}u_{\mathrm{ref},j},
\end{align}
for the zeroth order,
\begin{align}
    \label{eq:Galileaninvariance2}
    \overline{\mathcal{U}_i\mathcal{U}_j}-\mathcal{A}_{ij}^{\mathrm{mod}}(\overline{\mathcal{U}}_i) &= \overline{u_iu_j} - \mathcal{A}_{ij}^{\mathrm{mod}}(\overline{u}_i) \nonumber\\&+ u_{\mathrm{ref},i}\underbrace{\left(\overline{u}_j-\overline{\overline{u}}_j + \dfrac{\sigma^2}{2}\dfrac{\partial^2\overline{\overline{u}}_j}{\partial x_k \partial x_k}\right)}_{\mathcal{O}(\sigma^4)} + \underbrace{\left(\overline{u}_i-\overline{\overline{u}}_i + \dfrac{\sigma^2}{2}\dfrac{\partial^2\overline{\overline{u}}_i}{\partial x_k \partial x_k}\right)}_{\mathcal{O}(\sigma^4)}u_{\mathrm{ref},j},
\end{align}
for the second order, and 
\begin{align}
    \label{eq:Galileaninvariance4}
    \overline{\mathcal{U}_i\mathcal{U}_j}-\mathcal{A}_{ij}^{\mathrm{mod}}(\overline{\mathcal{U}}_i) &= \overline{u_iu_j} - \mathcal{A}_{ij}^{\mathrm{mod}}(\overline{u}_i) \nonumber\\&+ u_{\mathrm{ref},i}\underbrace{\left(\overline{u}_j-\overline{\overline{u}}_j + \dfrac{\sigma^2}{2}\dfrac{\partial^2\overline{\overline{u}}_j}{\partial x_k \partial x_k} - \dfrac{\sigma^4}{4}\dfrac{\partial^4\overline{\overline{u}}_j}{\partial x_l \partial x_l\partial x_k \partial x_k}\right)}_{\mathcal{O}(\sigma^6)} \nonumber\\&+ \underbrace{\left(\overline{u}_i-\overline{\overline{u}}_i + \dfrac{\sigma^2}{2}\dfrac{\partial^2\overline{\overline{u}}_i}{\partial x_k \partial x_k} - \dfrac{\sigma^4}{4}\dfrac{\partial^4\overline{\overline{u}}_i}{\partial x_l \partial x_l \partial x_k \partial x_k}\right)}_{\mathcal{O}(\sigma^6)}u_{\mathrm{ref},j},
\end{align}
for the fourth order. Since additional terms arise, Galilean invariance is not satisfied exactly. However, the third and fourth terms on the right-hand sides of equations \eqref{eq:Galileaninvariance0}-\eqref{eq:Galileaninvariance4} are small in the sense that they converge to zero for small filter widths with different orders. By comparing the terms with the Taylor series \eqref{eq:Tayolorseriesphi}, it can be seen that they are of the same order in $\sigma$ as the approximation for the filtered advection term. Therefore, as the order in $\sigma$ increases, Galilean invariance is satisfied with increasing accuracy. This conclusion is not of a different quality than the violation of Galilean invariance of the commonly applied flux limiters and stabilization terms. The quantitative effect is studied in section \ref{ssec:Galileaninvariance}.

\section{A posteriori studies}
\label{sec:aposteriori}
\subsection{Decaying turbulence}
\label{ssec:decayingturbulence}
\subsubsection{Simulation setup}
As a first a posteriori test case, decaying turbulence in a fully periodic cubic domain is considered. The decay starts from the initially laminar and divergence-free velocity field 
\begin{align}
\boldsymbol{u} &= U_0\begin{pmatrix} \sin{(2\pi y/L)} \\ \sin{(2\pi z/L)} \\ \sin{(2\pi x/L)}\end{pmatrix},
\end{align}
where $L$ is the size of the cubic domain and $U_0$ the maximum velocity magnitude in each direction. The filtered initial velocity field with the filter width $\sigma$ is therefore
\begin{align}
\overline{\boldsymbol{u}} &= U_0\begin{pmatrix} \sin{(2\pi y/L)} \\ \sin{(2\pi z/L)} \\ \sin{(2\pi x/L)}\end{pmatrix}\exp{(-(2\pi \sigma)^2/2L^2)}.
\end{align}
The Reynolds number can be defined as $\mathrm{Re}= \rho_\mathrm{f}U_0 L/\mu_\mathrm{f}$. In the present study, the Reynolds number is chosen as $\mathrm{Re}=1000$. After a relatively short laminar initial phase, the flow field rapidly turns turbulent and decays until the initial kinetic energy is fully dissipated. \\
\reviewera{The present configuration possesses well-defined initial conditions and requires no artificial forcing of turbulence. It is also a challenging case for LES, as it involves laminar-turbulent transition and a dynamically evolving kinetic energy spectrum.} \\
A DNS of the decaying turbulence is performed with a second order accurate finite volume solver that has been described and validated in many previous studies \citep{Denner2014b,Bartholomew2018,Denner2020,Hausmann2022a,Hausmann2024b}. The DNS are solved on a spatially uniform mesh with $N^3=256^3$ cells and a CFL-number of $0.1$. \\
The LES are performed with the same finite volume solver as the DNS but with $N^3=64^3$ mesh cells and with a filter width $\sigma=\Delta x = L/N$, if not stated otherwise. The LES cases differ in the representation of the filtered advection term and if the model for the filtered advection term is supplemented with a turbulent viscosity according to the Smagorinsky model or not. The filtered advection term is approximated with the proposed second order or fourth order expression derived in section \ref{ssec:defiltering}. For comparison, classical LES are performed that use either central differences or the van Leer flux limiter for the advection term. Details on the implementation of the central differences and the van Leer flux limiter in the present finite volume solver can be found in \citet{Denner2015}. 

\subsubsection{Results}
\label{ssec:results}
\begin{figure}
    \centering
    \includegraphics[width=1\linewidth]{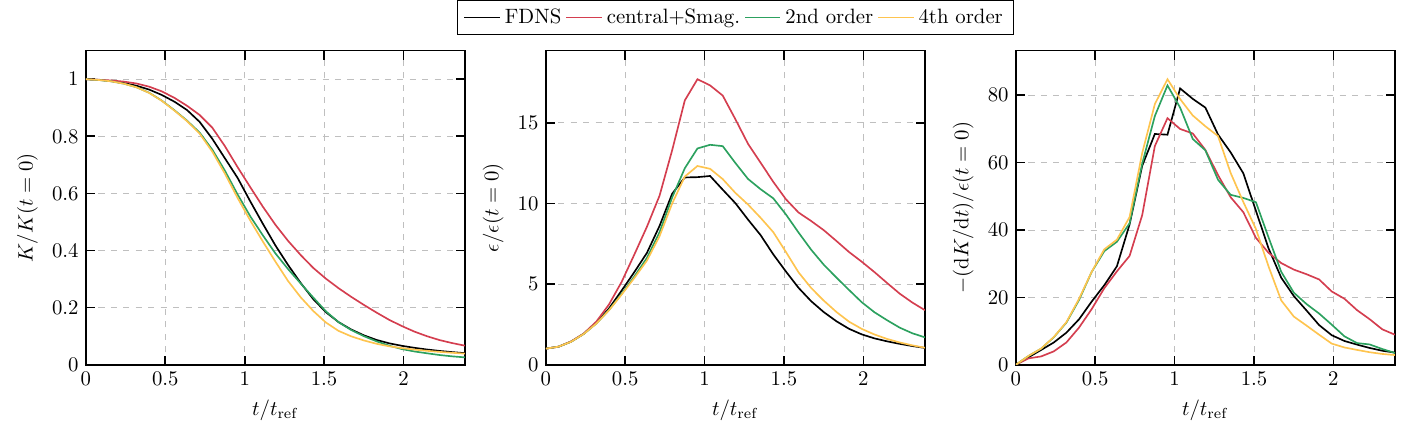}
    \caption{Temporal evolution of the filtered kinetic energy, viscous dissipation, and the total energy removal for the FDNS, the classical LES with central differences and the Smagorinsky model, and the LES with the second and fourth order approximation of the filtered advection term.}
    \label{fig:decaystatistics}
\end{figure}

\begin{figure}
    \centering
    \includegraphics[width=1\linewidth]{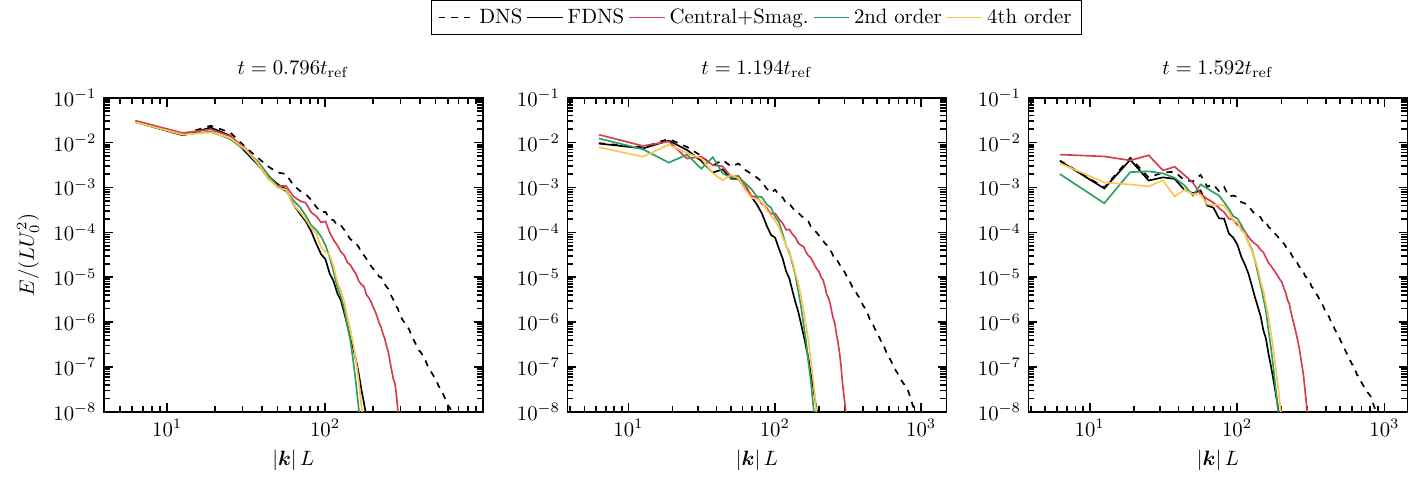}
    \caption{Kinetic energy spectrum at different times during the decay for the FDNS, the classical LES with central differences and the Smagorinsky model, and the LES with the second and fourth order approximation of the filtered advection term. The DNS spectrum is shown as reference. }
    \label{fig:spectra}
\end{figure}

\begin{figure}
    \centering
    \includegraphics[width=1\linewidth]{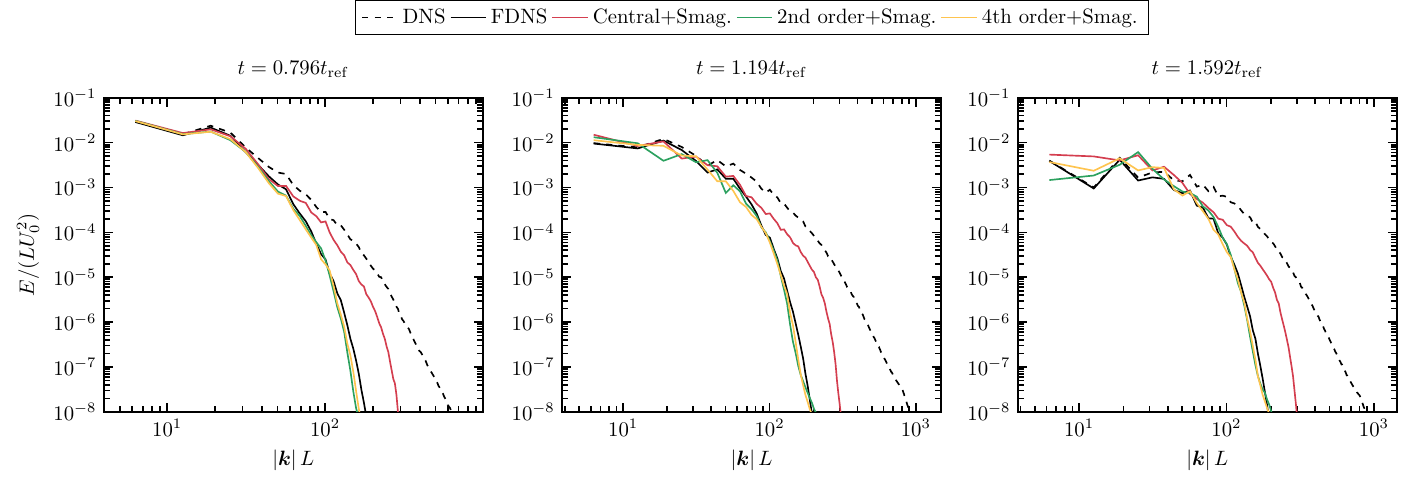}
    \caption{Kinetic energy spectrum at different times during the decay for the FDNS, the classical LES with central differences and the Smagorinsky model, and the LES with the second and fourth order approximation of the filtered advection term with the Smagorinsky model. The DNS spectrum is shown as reference.}
    \label{fig:spectraSmag}
\end{figure}

\begin{figure}
    \centering
    \includegraphics[width=1\linewidth]{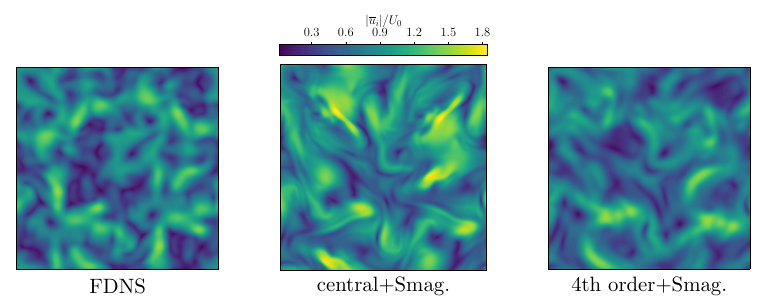}
    \caption{Slice of the absolute velocity during the decay at $t=1.194t_\mathrm{ref}$ for the FDNS, the classical LES with central differences and the Smagorinsky model, and the fourth order approximation of the filtered advection term with the Smagorinsky model.}
    \label{fig:contoursdecay}
\end{figure}

In figure \ref{fig:decaystatistics}, the temporal evolution of different flow statistics is shown. In particular, figure \ref{fig:decaystatistics} shows the kinetic energy of the filtered velocity
\begin{align}
    K = \dfrac{1}{2}\langle\overline{u}_i\overline{u}_i\rangle,
\end{align}
the viscous dissipation of the filtered velocity field
\begin{align}
    \epsilon = \mu_\mathrm{f} \langle\dfrac{\partial \overline{u}_i}{\partial x_j}\dfrac{\partial \overline{u}_i}{\partial x_j}\rangle,
\end{align}
and the total removal of energy from the filtered scales that includes the energy transferred to the subfilter scales and that is given as $-\mathrm{d}K/\mathrm{d}t$. The kinetic energy of the filtered scales is normalized by its initial value $K(t=0)$ and the viscous dissipation and the total energy removal are normalized by the initial viscous dissipation $\epsilon(t=0)$. The time is normalized by a reference time $t_\mathrm{ref}=L/U_0$. \\
The temporal evolution of the filtered kinetic energy predicted by the LES with the second and fourth order approximation of the filtered advection term closely resembles that of the FDNS. The classical LES with central differences and the Smagorinsky model decays too slow. Both observations are consistent with the a priori investigations of the energy transfer spectrum in section \ref{ssec:energytransfer} that show that the dyadic product of the filtered velocities with the Smagorinsky model removes too little energy from the large filtered scales and transfers too much energy to the small filtered scales. Furthermore, it can be observed in figure \ref{fig:decaystatistics} that the viscous dissipation of the filtered field is overpredicted by the classical LES with the Smagorinsky model. This seems to contradict the too slow energy decay of the classical LES. However, the viscous dissipation of the filtered field makes up only approximately $15\%$ of the total energy removal at the present filter width. Therefore, the majority of the energy removal from the filtered scales is caused by the filtered advection term. The second and fourth order approximation of the filtered advection term predict the viscous dissipation of the filtered field and the total energy removal well, whereas including terms of higher order leads to improved results. \\
More insights can be obtained by considering the kinetic energy spectrum at different times of the decay as shown in figure \ref{fig:spectra}. At all of the three time instants, the classical LES with the Smagorinsky model predicts too high energy at the high wave numbers, confirming the insufficient removal of energy. The approximations of the filtered advection term including terms up to second and fourth order predict a much more accurate spectrum, but the energy at the smallest filtered scales is slightly overestimated consistent with the a priori observation that only a part of the energy transfer is captured that increases with the order. It should be noted that the spectra at the small wave numbers are not statistically meaningful because single realizations are considered and only very few small wave number modes exist. \\
As suggested by the a priori analysis, supplementing the second and fourth order approximation of the filtered advection term with a turbulent viscosity according to the Smagorinsky model can give the right total energy transfer and even the right spectral distribution of the energy transfer. In figure \ref{fig:spectraSmag} the same kinetic energy spectra are shown but the second and fourth order approximations of the filtered advection term are supplemented with the Smagorinsky model. With the additional turbulent viscosity, the LES with the second and fourth order approximation of the filtered advection term reproduce the kinetic energy spectrum of the FDNS accurately for all the three time instants. \\
A known property of classical LES with a turbulent viscosity model is that they fail to predict the correct topology of flow structures of the filtered velocity and, instead, predict a flow field that is topologically similar to a turbulent flow with a smaller $\mathrm{Re}$ \citep{Kamal2024}. This is confirmed in figure \ref{fig:contoursdecay}, where the absolute filtered velocity is depicted in a slice during the turbulent decay for the FDNS, the classical LES with the Smagorinsky model, and the fourth order approximation of the filtered advection with the Smagorinsky model. The classical LES produces very elongated flow structures whereas the topology of the explicitly filtered velocity is much more isotropic. With the fourth order approximation of the filtered advection term, the filtered velocity structures are similar in size and magnitude to the FNDS. 

\subsubsection{Behavior under mesh refinement}
\label{ssec:meshrefinement}
\begin{figure}
    \centering
    \includegraphics[width=1\linewidth]{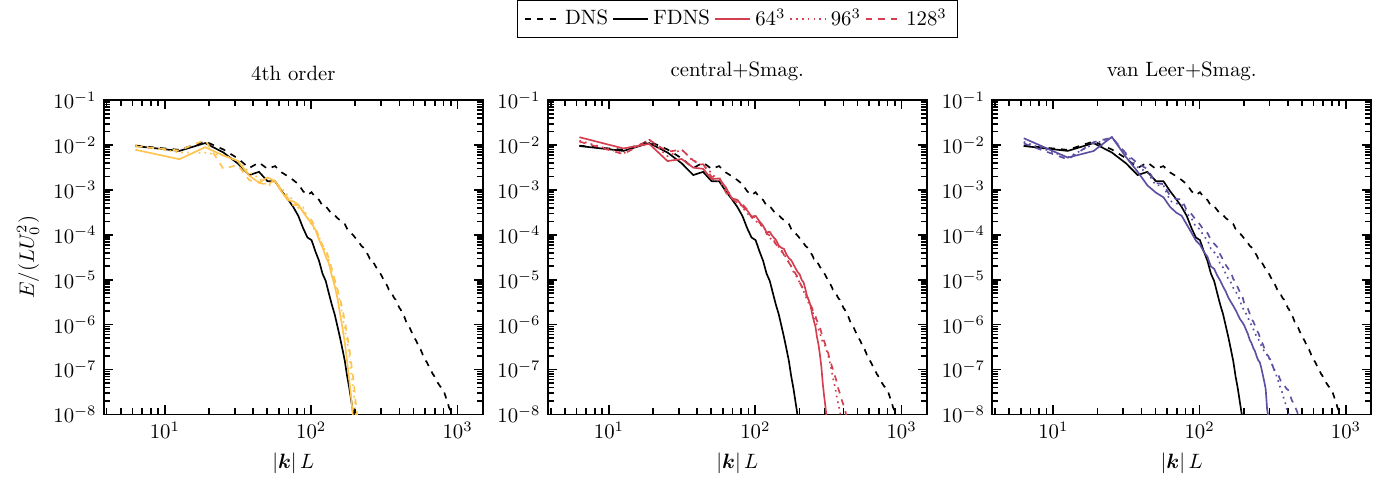}
    \caption{Kinetic energy spectrum at $t=1.194t_\mathrm{ref}$ and for different resolutions obtained with the fourth order approximation of the filtered advection term, the classical LES with the central differences and the Smagorinsky model, and the classical LES with the van Leer flux limiter and the Smagorinsky model.}
    \label{fig:spectraconvergence}
\end{figure}

A fundamental property when solving for the filtered flow field is that the solution converges under mesh refinement, i.e., after a minimum resolution is ensured, further refinement does not alter the statistics of the filtered flow field. Figure \ref{fig:spectraconvergence} shows the kinetic energy spectra at $t=1.194t_\mathrm{ref}$ of the LES with the fourth order approximation of the filtered advection term, the classical LES with the central differences and the Smagorinsky model, and the classical LES with the van Leer flux limiter and the Smagorinsky model with $64^3$, $96^3$, and $128^3$ mesh cells. The kinetic energy spectrum with the fourth order approximation is essentially independent of the resolutions, which means that $\Delta x = \sigma$ is a sufficient resolution for the fourth order approximation. With the classical LES and central differences, the kinetic energy spectrum possesses much larger wave numbers and the spectrum changes as the resolution is increased from $64^3$ to $96^3$. Further refinement does not influence the spectrum because the turbulent viscosity from the Smagorinsky model dissipates the very small flow structures. The classical LES with the van Leer flux limiter behaves differently. The kinetic energy spectrum differs significantly from the spectrum obtained with central differences and no convergence is observed under the tested refinements. This shows that the statistics of classical LES can strongly depend on the advection scheme and the resolution. 

\subsubsection{Galilean invariance}
\label{ssec:Galileaninvariance}
\begin{figure}
    \centering
    \includegraphics[width=1\linewidth]{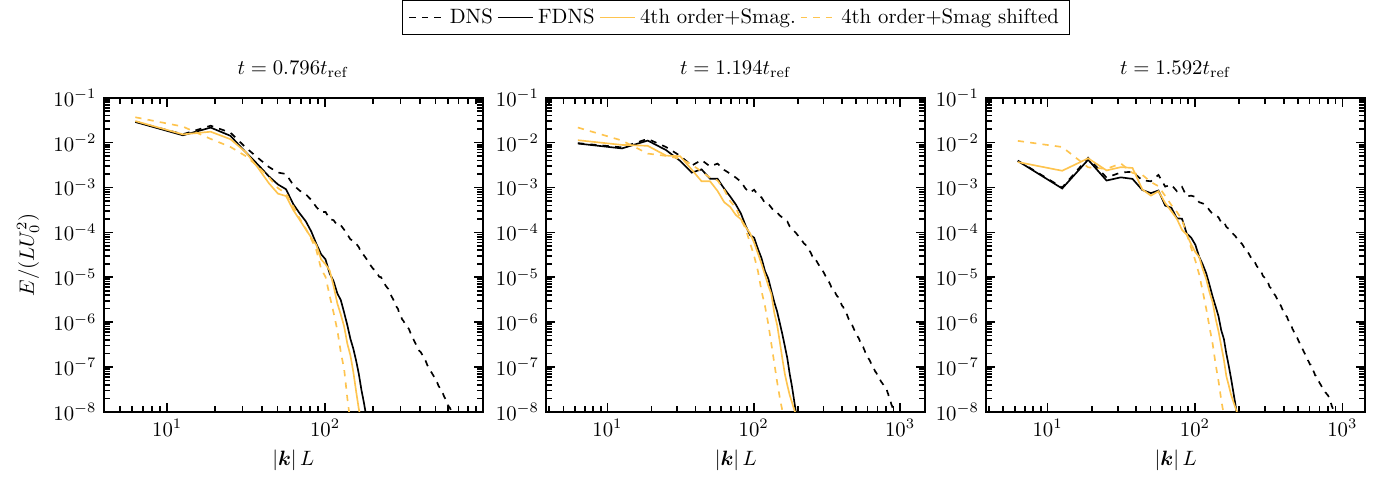}
    \caption{Kinetic energy spectrum at different times during the decay for the FDNS and the LES with the 4th order approximation of the filtered advection term with a turbulent viscosity according to the Smagorinsky model performed in different frames of reference.}
    \label{fig:spectrashifted}
\end{figure}

A potential drawback of the proposed approximation of the filtered advection term is that it is not exactly Galilean invariant, but only so up to the order of the accuracy of the truncation. To assess the practical implications, the LES of the fourth order approximation with the Smagorinsky model are repeated in a frame of reference shifted by $10U_0$, i.e., by ten times the amplitude of the initial velocity field. An exactly Galilean invariant model should predict the identical statistics after the results are transformed back into the original frame, which is the case for the central differences but not for the van Leer flux limiter or other common flux limiters. Figure \ref{fig:spectrashifted} shows the kinetic energy spectra of the results in the shifted frame of reference after transformation back into the original frame compared to the spectra obtained with the LES in the original frame. It is observed that the spectra are indeed not identical, but the differences are minor and of the order of the uncertainty of the approximation itself. Since even a significant change of the frame of reference causes only a minor change in the turbulence statistics, the approximate fulfillment of Galilean invariance does not seem to be of practical relevance for most of the cases. 

\subsection{Turbulent shear flow}

\subsubsection{Simulation setup}
\begin{figure}
    \centering
    \includegraphics[width=0.5\linewidth]{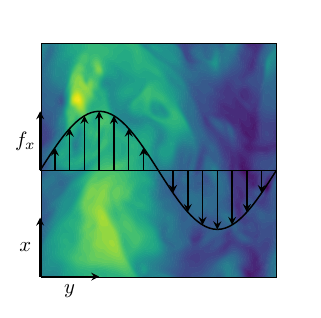}
    \caption{Illustration of the turbulent shear flow configuration that is driven by a momentum source $f_x$ in the x-direction varying with the shape of a sine in the y-direction.}
    \label{fig:sketchshearflow}
\end{figure}

A statistically inhomogeneous and anisotropic turbulent shear flow is considered as second test case to evaluate the proposed approximations for the filtered advection term. The same turbulent shear flow has been investigated in previous studies \citep{Hausmann2022a,Hausmann2023a} and is briefly summarized in the following. \\
The turbulent shear flow is solved in a fully periodic cube with the side length $L$ and the it is driven by a  sine-shaped momentum source $\boldsymbol{f}$, as illustrated in figure \ref{fig:sketchshearflow}. Mathematically, the momentum source driving the shear flow is given as
\begin{align}
    \boldsymbol{f}(y) = f_{\mathrm{max}} \sin (2\pi y/L)\boldsymbol{e}_x,
\end{align}
where $f_{\mathrm{max}}$ is the amplitude of the momentum source and $\boldsymbol{e}_x$ is the unit vector in the x-direction. With these quantities, a velocity scale $u_\mathrm{shear}$ and a corresponding Reynolds number $\mathrm{Re_{shear}}=\rho_\mathrm{f}u_\mathrm{shear}L/\mu_\mathrm{f}$ can be defined that is $\mathrm{Re_{shear}}=3115$ in the present case. The DNS, which is described in more detail in \citet{Hausmann2022a}, is carried out with the same finite volume solver that is used for the decaying turbulence on a uniform mesh with $N^3=128^3$ cells. \\
LES are performed on coarser meshes with $N^3=32^3$ and $N^3=16^3$ cells and are driven by the filtered momentum source 
\begin{align}
    \overline{\boldsymbol{f}}(y) = f_{\mathrm{max}} \sin (2\pi y/L)\exp{(-(2\pi \sigma)^2/2L^2)}\boldsymbol{e}_x.
\end{align}
\reviewera{The classical LES are carried out with the Smagorinsky model with a constant $0.2$ and with the Vreman model with a constant $0.1$. Latter has been specifically derived for shear dominated turbulent flows \citep{Vreman2004}.}

\subsubsection{Results}
\begin{figure}
    \centering
    \includegraphics[width=0.85\linewidth]{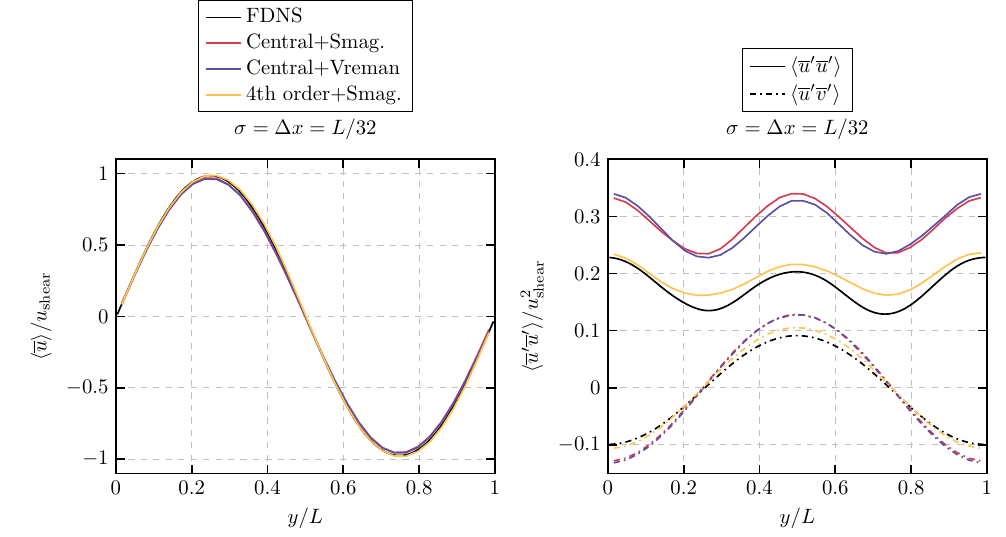}
    \includegraphics[width=0.85\linewidth]{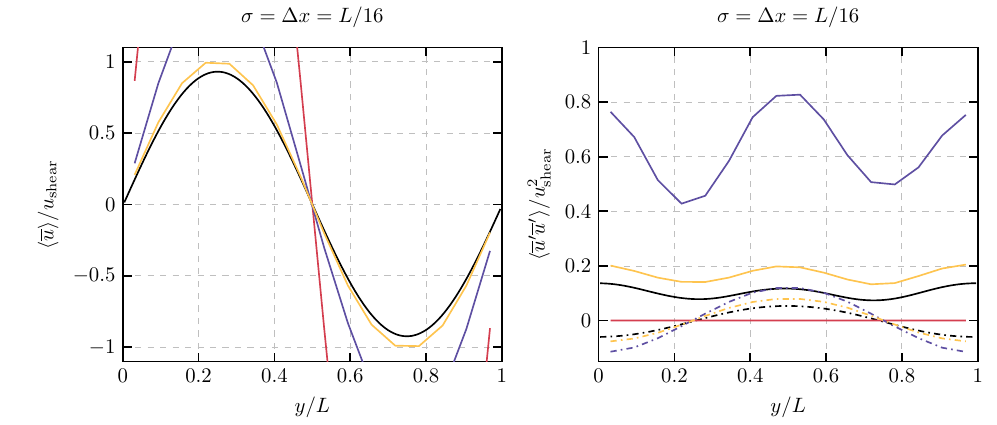}
    \caption{Filtered mean velocity and correlations of the filtered velocity fluctuations for two different filter widths of the turbulent shear flow configuration. The results of the classical LES with central differences and LES with the fourth order series expansion are shown, both with the Smagorinsky model. \reviewera{LES results with central differences and the Vreman model are also shown.}}
    \label{fig:tsprofiles}
\end{figure}

Figure \ref{fig:tsprofiles} shows the mean of the filtered velocity and the mean correlations of the fluctuations of the filtered velocity for the different LES, where a fluctuation of a flow quantity $\varPhi$ is indicated as $\varPhi^\prime=\varPhi - \langle \varPhi\rangle$. 
\reviewera{With the finer resolution and the small filter width $\sigma=\Delta x = L/32$, all of the LES simulations (central differences with Smagorinsky, central differences with Vreman, and fourth order approximation with Smagorinsky) predict the mean filtered velocity profile very accurately}. However, differences are observed in the filtered velocity correlations. The classical LES with central differences predict a too large magnitude of $\langle\overline{u}^\prime\overline{v}^\prime\rangle$ and significantly overpredict $\langle\overline{u}^\prime\overline{u}^\prime\rangle$, whereas the fourth order approximation of the advection term leads to accurate predictions of the correlations. The significant overprediction of the streamwise correlations by classical LES is a well-known phenomenon from wall-bounded flows that can be, at least partially, attributed to the inconsistent treatment of wall-boundary conditions in coarse LES (see, e.g., \citet{Bae2018}). The fact that $\langle\overline{u}^\prime\overline{u}^\prime\rangle$ is overestimated by classical LES in an unconfined turbulent shear flow suggests that the misalignment of the dyadic product of the filtered velocities with the filtered advection term also plays a role. With the fourth order approximation that is shown to be highly correlated with the filtered advection term, $\langle\overline{u}^\prime\overline{u}^\prime\rangle$ is predicted well. \\
Figure \ref{fig:tsprofiles} also shows the results for coarser LES with $\sigma=\Delta x = L/16$. \reviewera{With such a coarse resolution, the classical LES with the Smagorinsky model does not turn turbulent and even relaminarizes when being initialized from a turbulent field which is why it leads to a very large mean velocity and zero fluctuations. The classical LES with the Vreman model leads to a turbulent flow with a singificant overestimation of the mean velocity and velocity fluctuations.} The explicitly filtered mean velocity differs only slightly from the filtered mean velocity with $\sigma=\Delta x = L/32$ and is predicted accurately with the LES using the fourth order approximation of the filtered advection term. The correlations of the filtered velocity fluctuations change drastically with the increase in filter width, which is captured to some extent by the LES with the fourth order approximation of the filtered advection term.

\section{Guidelines for implementation}
\label{sec:implementation}
Although the approximation for the filtered advection term including terms up the fourth order, as given in equation \eqref{eq:fourthorder}, is a relatively long expression, its implementation is conceptually simple. It is very common that flow solvers that rely on the dyadic product of the filtered velocities as advection term treat the advection term at least partially implicit. Typically, iterations are carried out during which the non-linear advection term is updated with filtered velocities from the previous iteration until a norm of the difference of the predicted filtered velocities between two consecutive iterations is below a threshold. When the proposed approximation of the filtered advection term is implemented, the implementation of the classical advection term (dyadic product of the filtered velocities) is replaced with a discretized form of equation \eqref{eq:zerothorder}, \eqref{eq:secondorder}, or \eqref{eq:fourthorder}, depending on the desired order. The structure of the iterations over the non-linear term depends on the specific flow solver but in general one iteration can be summarized as follows:
\begin{enumerate}
    \item Compute the approximation of $u_i u_j$ (without the filtering operation) from the filtered velocity of the previous iteration up to the desired order given in equations \eqref{eq:zerothorder}-\eqref{eq:fourthorder}.
    \item Perform the explicit filtering operation with a Gaussian filter kernel to obtain an approximation of $\overline{u_i u_j}$ from the previously computed approximation of $u_i u_j$.
    \item Solve the filtered NSE with the approximation for $\overline{u_i u_j}$ as the advection term to obtain the predicted filtered velocity of the present iteration.
    \item Compute a norm of the difference between the predicted velocity of the present iteration and the predicted velocity of the previous iteration.
    \item If the norm exceeds the convergence threshold, start another iteration. Otherwise, start with the next time step. 
\end{enumerate}
This solution algorithm proves to be numerically very robust. For instance, even when repeating the simulations of the decaying turbulence configuration with the fourth order approximation of the filtered advection term as discussed in section \ref{ssec:decayingturbulence} but with zero viscosity (infinite Reynolds number) and a CFL-number of one, a stable numerical solution is obtained requiring not more than three to four iterations per time step.

\section{Conclusions}
\label{sec:conclusions}

This article addresses the widely overlooked conceptual inconsistency of classical LES that the commonly used advection term, $\overline{u}_i\overline{u}_j$, introduces higher wave numbers in the filtered NSE than consistent with the definition of a filtered equation. Consequently, the filtered velocity gets contaminated with high wave number content and numerical interventions are required, such as flux limiters, stabilization terms, or dealiasing, to capture the solution with coarse numerical resolution. Alternatively, the filtered advection term is modeled directly in the present article with an expression that contains only filtered terms and, therefore, leads to consistent filtered NSE. An expression is derived that is an infinite series expansion in powers of the filter width. As shown in a priori studies of HIT, retaining only a few terms of the series expansion leads to a model that is strongly correlated with the filtered advection term. A posteriori studies of decaying turbulence and a turbulent shear flow demonstrate that the proposed approximation of the filtered advection term is stable without any additional numerical interventions and that it predicts realistic filtered flow structures and filtered velocity correlations. Furthermore, the predicted energy spectrum is shown to converge under mesh refinement and that Galilean invariance is satisfied to good approximation in practice. The consistent and accurate approximation of the filtered advection term proposed in the present article represents a step towards more reliable and interpretable LES.

\section*{Data Availability Statement}
The data that support the findings of this article are openly available in the repository https://doi.org/10.5281/zenodo.18999736. 
\begin{acknowledgments}
    This research was funded by the Deutsche Forschungsgemeinschaft (DFG, German Research Foundation): \textemdash Project-ID 457509672.
\end{acknowledgments}


%

\appendix
\section{Spectral decay of the filtered advection term}
\label{ap:spectraldecay}
Assuming that the velocity is bounded for all wave numbers, i.e., there exists a real finite constant $C_\mathrm{u}$ such that
\begin{align}
    |\hat{u}_i(\boldsymbol{k})|<C_\mathrm{u}, \quad \forall \boldsymbol{k}\in\mathbb{R}^3,
\end{align}
where $\hat{.}$ indicates the Fourier transform, the filtered advection term $\overline{u_i u_j}$ decays as rapidly as a Gaussian in spectral space
\begin{align}
    \lim_{|\boldsymbol{k}|\rightarrow\infty}|\hat{G}(\boldsymbol{k})\widehat{u_iu_j}(\boldsymbol{k})|< \lim_{|\boldsymbol{k}|\rightarrow\infty}|C_\mathrm{G}\hat{G}(\boldsymbol{k})|,
\end{align}
for a finite real constant $C_\mathrm{G}$.  \\
In the following it is shown that the proposed approximation of $\overline{u_i u_j}$, which contains terms of the type 
\begin{align}
    \overline{\dfrac{\partial^{n}\overline{u}_i\overline{u}_j}{\partial(\sigma^2)^{n}}}, \nonumber
\end{align}
also decays at least as rapidly as a Gaussian with standard deviation $\sigma$. \\
After applying the convolution theorem, the Fourier transform of the terms in the series expansion can be written as
\begin{align}
\widehat{
\overline{
\frac{\partial^n}{\partial(\sigma^2)^n}
\bigl(\overline{u}_i\,\overline{u}_j\bigr)
}}(\boldsymbol{k})
= \hat{G}(\boldsymbol{k})\,
\frac{\partial^n}{\partial(\sigma^2)^n}
\widehat{
\overline{u}_i\,\overline{u}_j
}(\boldsymbol{k}).
\end{align}
Differentiating the Fourier transform of a Gaussian with respect to $\sigma^2$ gives
\begin{align}
    \dfrac{\partial \hat{G}(\boldsymbol{k})}{\partial (\sigma^2)} = -\dfrac{|\boldsymbol{k}|^2}{2}\hat{G}(\boldsymbol{k}),
\end{align}
such that the $n$th derivative of the Fourier transform of the dyadic product is given as
\begin{align}
\frac{\partial^n}{\partial(\sigma^2)^n}
\widehat{
\overline{u}_i\,\overline{u}_j
}(\boldsymbol{k})
&=
\frac{\partial^n}{\partial(\sigma^2)^n}\int_{\mathbb{R}^3}
\hat{G}(\boldsymbol{p})\hat{G}(\boldsymbol{k}-\boldsymbol{p})\,
\hat{u}_i(\boldsymbol{p})\hat{u}_j(\boldsymbol{k}-\boldsymbol{p})
\,\mathrm{d}\boldsymbol{p} \nonumber\\
&=
\int_{\mathbb{R}^3}
P(|\boldsymbol{p}|^2,|\boldsymbol{k}-\boldsymbol{p}|^2)\,
\hat{G}(\boldsymbol{p})\hat{G}(\boldsymbol{k}-\boldsymbol{p})\,
\hat{u}_i(\boldsymbol{p})\hat{u}_j(\boldsymbol{k}-\boldsymbol{p})
\,\mathrm{d}\boldsymbol{p},
\end{align}
where $P$ is a polynomial.
Since the velocity is bounded, there exists a real finite constant $C_{ij}$ that satisfies
\begin{align}
\lim_{|\boldsymbol{k}|\to\infty}\left|
\widehat{
\overline{
\frac{\partial^n}{\partial(\sigma^2)^n}
\bigl(\overline{u}_i\,\overline{u}_j\bigr)
}}(\boldsymbol{k})
\right|
<
\lim_{|\boldsymbol{k}|\to\infty}\left|C_{ij}\hat{G}(\boldsymbol{k})
\int_{\mathbb{R}^3}
\left|P\right|
\hat{G}(\boldsymbol{p})\hat{G}(\boldsymbol{k}-\boldsymbol{p})
\,\hat{u}_i(\boldsymbol{p})\hat{u}_j(\boldsymbol{k}-\boldsymbol{p})
\,\mathrm{d}\boldsymbol{p}\right|.
\end{align}
Since a Gaussian decays faster than any polynomial can grow,
\begin{align}
    \int_{\mathbb{R}^3}
\left|P\right|
\hat{G}(\boldsymbol{p})\hat{G}(\boldsymbol{k}-\boldsymbol{p})
\,\hat{u}_i(\boldsymbol{p})\hat{u}_j(\boldsymbol{k}-\boldsymbol{p})
\,\mathrm{d}\boldsymbol{p} \nonumber
\end{align}
is bounded and there exists a real finite constant $C_\mathrm{G}$ such that
\begin{align}
\lim_{|\boldsymbol{k}|\to\infty}
\left|
\widehat{
\overline{
\frac{\partial^n}{\partial(\sigma^2)^n}
\bigl(\overline{u}_i\,\overline{u}_j\bigr)
}}(\boldsymbol{k})
\right|
<
\lim_{|\boldsymbol{k}|\to\infty}
\big|C_\mathrm{G} \hat{G}(\boldsymbol{k})\big|.
\end{align}


\end{document}